\documentclass{jfm}
\usepackage{graphicx}
\usepackage{xcolor}
\usepackage{epstopdf, epsfig}
\usepackage{multirow, booktabs}
\usepackage{amsfonts}
\usepackage{amsmath}
\usepackage{setspace} 
\usepackage{bm} 
\usepackage{enumitem}
\usepackage{xstring}
\usepackage{url}

\newcommand{\PP}{\tau}
\newcommand{\press}{p_L}

\usepackage{cleveref} 

\makeatletter
\def\tagform@#1{%
  \maketag@@@{(#1\unskip\@@italiccorr)}%
}
\makeatother

\crefname{equation}{equation}{equations}
\Crefname{equation}{Equation}{Equations}

\shorttitle{Variational approach to droplet motion}
\shortauthor{G.~I.~T\'oth, D.~N.~Sibley, A.~J.~Bok\'anyi-T\'oth, D.~Tseluiko \& A.~J.~Archer}

\title{Variational approach to droplet motion on uneven solid surfaces, including
contact line dynamics and evaporation}

\author{
Gyula~I.~T\'oth\corresp{\email{G.I.Toth@lboro.ac.uk}},
David~N.~Sibley\corresp{\email{D.N.Sibley@lboro.ac.uk}},
Agnes~J.~Bok\'anyi-T\'oth\corresp{\email{A.J.Bokanyi-Toth@lboro.ac.uk}},
Dmitri~Tseluiko\corresp{\email{D.Tseluiko@lboro.ac.uk}}
\and
Andrew~J.~Archer\corresp{\email{A.J.Archer@lboro.ac.uk}}
}

\affiliation{Department of Mathematical Sciences and Interdisciplinary Centre for Mathematical Modelling, Loughborough University,
Loughborough, LE11 3TU, UK}

\begin{document}

\maketitle

\begin{abstract}
We show how dynamical equations for liquid films and drops on uneven surfaces, including contact line dynamics and evaporation/condensation effects, may be formulated as a variational dynamics, generated via Onsager's variational principle. The theory applies in the isothermal overdamped-dynamics limit. We apply this general approach to obtain several well-known results on contact line dynamics and to study drops pinning and sliding on inclined corrugated surfaces. This approach constructs the dynamical equations starting from the free energy of the system and therefore has the advantage that it naturally incorporates the correct equilibrium properties.
\end{abstract}


\section{Introduction}

In mechanics, there is a long history of formulating the dynamical equations as a function(al) minimization problem.
For overdamped dissipative systems, the Onsager reciprocal relations and minimisation principle can be used to show that the dynamics may be expressed as a minimum free energy dissipation principle \citep{onsager1931reciprocal_1, groot1964non}.
More recently, Onsager's variational principle (OVP) has been used with great success to derive equations of motion for fluids and other soft condensed matter systems in the isothermal overdamped-dynamics limit, which includes the low-Reynolds-number regime.
An introduction to this approach, giving a good discussion of some of the key ideas, are the works of Doi et al.~\citep{doi2011onsager, doi2013soft, doi2019application, doi2021onsager} and also \cite{wang2021onsager}.
Here we apply OVP to generate the equations of motion for liquid films and drops on surfaces.
Other systems where OVP works well include Brownian dynamics (colloidal suspensions) \citep{schmidt2013power, schmidt2022power, archer2026variational}, soft matter phase separation (Cahn-Hilliard theory and its generalisations), polymer and gel dynamics, viscoelastic fluids \citep{doi2011onsager, doi2013soft}, liquid bridges \citep{bokanyi2026liquid} and active matter \citep{wang2021onsager}, to mention just a few examples.

The starting point for the OVP based approach is that one must have a reliable approximation for the free energy of the system $F$, expressed as a function/functional of the relevant slow variables/fields characterizing the state of the system.
In other words, one must have an expression for $F$ that incorporates all of the relevant physics for the correct description of the static equilibrium behaviour of the system.
Thus, the equilibrium limit must already be fully understood.
For example, for liquid drops on surfaces, which are the particular systems of interest here, in the overdamped limit, the relevant slow field is the thickness of the liquid $h({\bf r})$ over the surface, where ${\bf r}$ is the location on the surface \citep{oron1997long, degennes2003capillarity, craster2009dynamics}.
Another relevant field may e.g.\ be the local concentration of solute particles within the liquid film.
For each relevant variable/field there is an associated velocity/current.
Next, one constructs the Rayleighan
\begin{equation}
\label{Rayleigh} \mathcal{R} = \dot{F}+\Phi \,,
\end{equation}
where $\dot{F}$ is the time derivative of the free energy and $\Phi$ is the dissipation function/functional.
In general, $\Phi$ is not known exactly -- this can most easily be seen when considering the formulation for systems of interacting Brownian particles, known as power functional theory \citep{schmidt2013power, schmidt2022power}.
However, the leading order contributions have to be quadratic (positive definite) in the velocity/current(s), and in many cases using just this quadratic approximation is sufficient.
To determine the dynamical mobility or friction coefficients corresponding to each dissipation channel in this quadratic requires input from either experiment or dynamical theory.
Without such knowledge, one can still make some progress by treating these as constants and using the expected dominant dissipation coefficient to scale time in the resulting dynamical equations.
The dynamical equations are then given by minimising the Rayleighan $\mathcal{R}$ with respect to the relevant velocities/currents.
We expand on these ideas in much more detail below for the case of liquid droplets on uneven surfaces.
It should now be evident to the reader that this variational approach has the potential to be very powerful.
It enables us to build on the extensive knowledge built up from thermodynamics, statistical mechanics, Landau theory, symmetry principles and so forth that all lead to good approximations for the free energy $F$ and then obtain in a simple manner time evolution equations applicable in the overdamped (slow dynamics) limit.

As mentioned, the systems of interest here are liquid droplets on uneven surfaces.
When for example drops slide on an inclined surface, the dynamics is governed by a coupling of the internal fluid flow, interaction with the surface and the dynamics of the contact line, where the perimeter of the droplet meets the surface \citep{DussanV_Davis_1974,DussanReview,deGennes1985,BLAKE20061,oron1997long, degennes2003capillarity, bonn2009wetting, craster2009dynamics, savva2013droplet, mchale2025surfaces}. 
There are also a wide variety of interesting works for droplets on rough or chemically patterned surfaces (sources of contact angle hysteresis), sometimes additionally including evaporation
\citep{schwartz1985contact, robbins1987contact, joanny1990motion, brandon1996simulation, konnur2000instability, cubaud2001faceted, kondic2002flow, moosavi2009dynamics, savva2009two, Raj, savva2011contact_pt1, savva2011contact_pt2, savva2011dynamics, savva2012influence, varagnolo2013stick, savva2013droplet, GrovesSavva21, ewetola2021control, bisquert2024competing, vrionis2025efficient}.
Here, we use OVP to derive the equations of motion for a drop moving over an uneven surface and its contact line dynamics.
We also show how the effects of evaporation or condensation may additionally be incorporated into the variational framework.

The background and starting point for the present work is \cite{qian2006variational}, where the case of droplets moving over a flat surfaces and the associated contact line dynamics is derived via OVP.
Later studies build on this, showing how OVP approach can be applied to liquid mixtures \citep{xu2015variational}, to describe droplets sliding down inclined surfaces \citep{xu2016variational}, connecting to the Stokes flow equations \citep{peschka2018variational} and showing the connection between the formulation for conserved and/or non-conserved fields \citep{Thie2018csa}, which is what we build on here to describe evaporation.
The appendix of \cite{lopes2018multiple} provides an insightful discussion related to this.
These works show the connections between the manner in which viscous dissipation is introduced in OVP approach (see below) and how it arises when the system dynamics is obtained starting instead from the Navier-Stokes equations.
In the main body of this paper the manner in which we introduce dissipation comes from the perspective of nonequilibrium thermodynamics.
However, in Appendix~\ref{app:df} we additionally give details showing the connections to fluid mechanics.

The thin-film equation for the time evolution of $h$ is more commonly derived by performing a long-wave analysis of the Navier-Stokes equations for a liquid on a surface \citep{oron1997long, craster2009dynamics}.
In contrast, here we obtain the thin-film equation starting from the free energy $F$ applied together with OVP (albeit here we retain the full curvature terms from the free energy, rather than expanding to obtain a square-gradient free energy). 
The connection between these two approaches is the gradient dynamics formulation of thin-film hydrodynamics \citep{mitlin1993dewetting, thiele2010thin, thiele2011note}.
A useful coarse-grained extension of the general OVP-based approach is to parametrise the droplet profile $h({\bf r},t)$ with just a few parameters, such as the radius and maximum height.
With these approximations, a series of studies \citep{man2017vapor, wu2018multi, wu2019drying, wu2021contact, yang2021deposition} show how to include the effects of evaporation, solute deposition and contact line motion, in order to model phenomena such as the interactions between droplets due to evaporation and the coffee ring effect.

The main contributions of the present work are (i) to obtain dynamical equations for contact line motion over uneven surfaces and how this is related to the contact angle and (ii) to show how evaporation/condensation can also be included.
Contact line motion is now fairly well understood \citep{snoeijer2013moving, sui2014numerical}.
There are a number of subtle issues that mean modelling the dynamics is not always straight-forward.
One common approach is to sidestep the issue by using a precursor film model, that requires incorporating into the free energy the so-called binding potential (closely related to the disjoining pressure). This is often the way to include much of the relevant wetting behaviour physics \citep{bonn2009wetting, degennes2003capillarity}, such as capturing near-wall structure/packing effects seen in microscale models \citep{nold2014fluid, nold2015nanoscale,hughes2015liquid, hughes2017influence} as well as for bubbles, freezing or colloidal behaviour \citep{Zhang_Sibley_Tseluiko_Archer_2024,Yin_2019,Llombart20,sibley2021ice}.
One can make good arguments as to why precursor film models do make sense even for `dry' surfaces \citep{yin2017films}.
However, here we instead directly tackle this issue by formulating the problem explicitly with a contact line (where the film height $h$ terminates) and derive equations of motion for the moving contact line.
This requires introduction of a slip-length in our mobility function; we say more about this below.

Another way of modelling the dynamics of moving contact lines is via phase-field or diffuse-interface methods \citep{SEPPECHER1996977,YueFengPoF,JACQMIN_2000,DingSpelt07,sibley2013moving,sibley2013contact,sibley2013unifying}. There is a wealth of excellent literature using this approach, and it is in this body of work where the language and techniques of the present work are most commonly seen. For example, by starting from writing an appropriate free energy that is minimised. Another particularly pertinent example is the behaviour of the contact angle and its connection to what is termed unbalanced/uncompensated Young’s force.
As discussed in the review articles of \cite{sui2014numerical, snoeijer2013moving}, various physical mechanisms or modelling approaches lead to the result that the difference in the cosines of the dynamic and static (equilibrium) contact angles is a key factor in the contact line dissipation. Anticipating our result in \S~\ref{subsec:contact_angle}, we see this result arise naturally through OVP.

This paper is structured as follows: In \S~\ref{sec:II} we present the theory for the dynamics of liquid drops on uneven surfaces.
We start this in \S~\ref{subsec:OVP}, by showing how to apply OVP to obtain the coupled equations for the drop motion and contact line dynamics over the surface.
This is followed by a few remarks in \S~\ref{subsec:equilib} on the equilibrium limit.
In \S~\ref{subsec:free_energy} we derive an approximation for the free energy of the system, incorporating the effects of gravity and capillarity.
This is followed in \S~\ref{subsec:contact_angle} with an analysis of the contact line dynamics, including deriving an equation for the contact line velocity in terms of the dynamic contact angle.
In \S~\ref{sec:evap} we show how to extend the theory of the previous section to include also the effects of evaporation or condensation.
In \S~\ref{sec:examples} we apply the theory of \S~\ref{sec:II} to four different specific situations, in order to illustrate what the theory can do and to demonstrate how it reproduces some known results.
These cases are: (\S~\ref{sec:liq_ridge}) a small liquid ridge (two-dimensional drop) spreading on a flat homogenous surface, (\S~\ref{subsec:inclined}) a liquid ridge sliding down an inclined corrugated surface, (\S~\ref{subsec:axisymmetric}) the case of an axisymmetric droplet spreading on an axisymmetrically corrugated surface and (\S~\ref{subsec:evap_example}) an evaporating or condensing liquid ridge.
Finally, in \S~\ref{sec:conc} we make a few concluding remarks.
The paper also includes four appendices, the first about the dissipation functional, the second gives some background on applying the calculus of variations when there are moving boundaries, the third about the dynamic contact line condition and the fourth gives the details of the numerical methods used to calculate the results in \S~\ref{sec:examples}.

\section{Theory for the dynamics}
\label{sec:II}

\subsection{Onsager variational approach}
\label{subsec:OVP}

\begin{figure}
    \centering
    \includegraphics[width=0.8\linewidth]{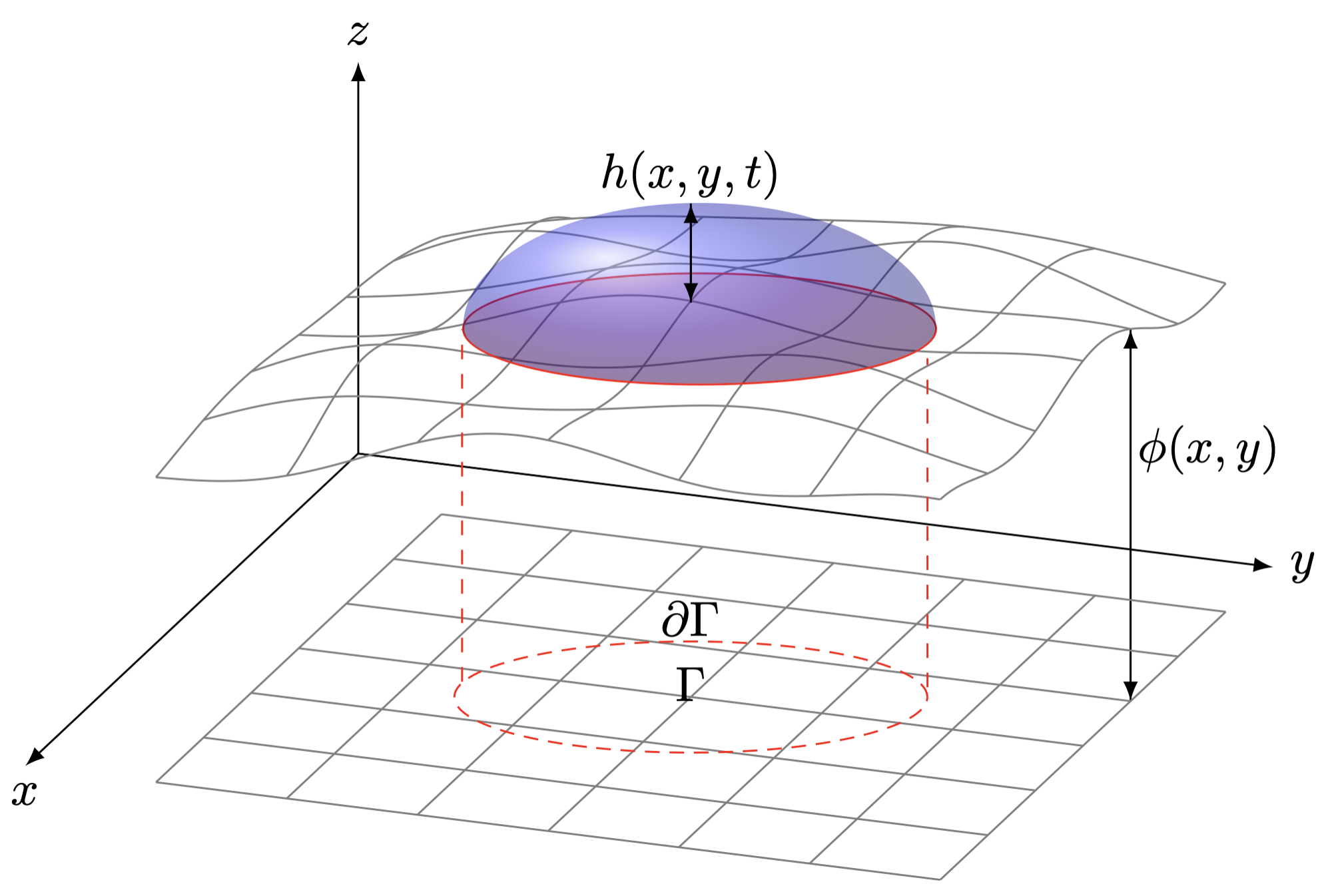}
    \caption{Sketch of a liquid droplet on an uneven solid surface. The solid surface has height profile $\phi(x,y)$. The liquid-gas interface is determined by the function $h(x,y,t)$, which gives the distance from the solid surface to the liquid-gas interface in the $z$ direction. The contact area on the surface is shaded red and its projection onto the $xy$-plane is denoted $\Gamma(t)$, while the projection of the contact line (where $h=0$) is the closed loop $\partial\Gamma(t)$.}
    \label{fig:drop_sketch}
\end{figure}

We consider a domain $V_0$ containing a solid object with a surface that is not necessarily flat and with a liquid drop sitting on the surface of the solid, as illustrated in Fig.~\ref{fig:drop_sketch}.
We choose $V_0$ to be sufficiently large that its boundaries are well away from the edge of the drop.
We assume also there is a gas (vapour phase) surrounding the liquid.
Let $\phi(\mathbf{r})$ denote the height of the solid surface measured from the $z=0$ plane, where $\mathbf{r}=(x,y)$, and let $h(\mathbf{r},t)$ be the height of the liquid-gas interface (droplet surface) above the surface of the solid. Let $\Gamma(t) \subset \mathbb{R}^2$ denote the support of $h(\mathbf{r},t)$, and let $\partial\Gamma(t)$ denote the boundary of $\Gamma(t)$.
Thus, $\partial\Gamma(t)$ is the projection onto the $z=0$ plane of the three-phase contact line around the edge of the drop. For the sake of simplicity, we assume that $\Gamma(t)$ is simply connected, and therefore $\partial\Gamma(t)$ is a simple plane curve. In our framework, $h(\mathbf{r},t)$ \textit{terminates} at the boundary $\partial\Gamma(t)$ and satisfies the Dirichlet boundary condition
\begin{equation}
\label{TBC} h(\mathbf{r},t)|_{\partial\Gamma(t)}=0 \,,
\end{equation}
where the notation $|_{\partial\Gamma}$ stands for ``evaluated at the domain boundary''.
The development of general dynamical equations for $h(\mathbf{r},t)$ and $\Gamma(t)$ starts with analysing the continuity equation. Assuming that the solid, liquid, and gas densities (denoted by $\rho_{S}$, $\rho_{L}$ and $\rho_{G}$, respectively) are constants, and that the flow is incompressible, the continuity equation can be written as $\Delta\dot{\rho} + \nabla \cdot (\Delta\rho\,\mathbf{v}) = 0$, where $\Delta\rho(\mathbf{x},t) = \rho(\mathbf{x},t) - \rho_G$ is the 3-dimensional (i.e., $\mathbf{x}=(x,y,z)$) mass density field relative to the gas density, and $\mathbf{v}(\mathbf{x},t)$ is the velocity field.
Note our use here of the dot over the variable to denote a (partial) time derivative.
We denote the position of the liquid-gas interface as
\begin{equation}\label{eq:xi}
\xi(x,y,t) \equiv \phi(x,y)+h(x,y,t).
\end{equation}
Assuming sharp interfaces, the mass density can be approximated as follows:
\begin{equation}\label{eq:rho_cases}
	\rho(x,y,z,t)\approx\begin{cases}
		\rho_S \quad\text{for}\quad z<\phi(x,y),\\
		\rho_L \quad\text{for}\quad \phi(x,y) \leq z< \xi(x,y,t),\\
		\rho_G \quad\text{for}\quad z\geq\xi(x,y,t).
	\end{cases}
\end{equation}
Assuming zero velocity field at the boundaries of the system domain $V_0$, integrating the continuity equation for the mass density with respect to $z$ and then dividing the result by $\Delta\rho_0 \equiv \rho_L-\rho_G$, results in the following governing dynamics for $h(\mathbf{r},t)$: 
\begin{equation}
\label{hdyng}
\dot{h} =  -\nabla \cdot \mathbf{J} \,,
\end{equation}
where $\nabla\cdot$ is the 2-dimensional divergence operator and
\begin{equation}
   \label{eq:J_def}
    \mathbf{J}(\mathbf{r},t) {=} \int_{\phi}^{\xi} 
    \,\mathbf{v}\,dz 
\end{equation}
is the flux, with $\mathbf{r}=(x,y) \in \Gamma(t)$.
In the absence of evaporation, condensation or any other phase transitions, the volume of the liquid
\begin{equation}\label{eq:volume}
    V[h,\Gamma] = \int_\Gamma h\,dA
\end{equation}
remains constant over time -- this is taken into account by Eq.~(\ref{hdyng}) and its boundary conditions.
Below in \S~\ref{sec:evap} we extend the present argument to include evaporation/condensation, but for simplicity we omit these effects for now.
The time derivative $\dot{V}$ can be calculated by applying the extension of the calculus of variations for varying domains (often referred to as the `transversality condition'), which yields (see Appendix~\ref{app:cv} for further details):
\begin{eqnarray}
\nonumber \dot{V} &=& \int_\Gamma \left\{ \dot{h}\,\frac{\delta V}{\delta h} \right\}\,dA + \int_0^1 \left\{\dot{\mathbf{R}} \cdot \left( h\,\mathbb{I} \right)_{\partial\Gamma} \cdot \mathbf{n}_\perp\right\}\,d\chi\\
\label{eq:7}
&=& -\int_{\Gamma} (\nabla\cdot\mathbf{J})\,dA = -\oint_{\partial\Gamma} \mathbf{J}\cdot d\mathbf{l}_\perp \,,
\end{eqnarray}
where we used Eq.~\eqref{hdyng}, $\frac{\delta V}{\delta h}=1$, $h|_{\partial\Gamma}=0$, and the divergence theorem. Furthermore, $\mathbf{R}(\chi,t)=(X(\chi,t),Y(\chi,t))$ stands for a parametric form of the domain boundary, where $\chi \in [0,1]$, $\mathbf{R}(0,t)=\mathbf{R}(1,t)$ (closed curve) for every $t \geq 0$, and the parametrisation satisfies the right-hand rule---i.e.\ increasing $\chi$ corresponds to the counter-clockwise direction along the curve.
The vector $\mathbf{n}_\perp =\left( Y_\chi(\chi,t), - X_\chi(\chi,t)\right)$ is outward normal to $\mathbf{R}(\chi,t)$, while $d\mathbf{l}_\perp=\mathbf{n}_\perp\,d\chi$. In the absence of phase transitions, $\dot{V}=0$, and therefore the flux must vanish on the boundary $\partial\Gamma(t)$:
\begin{equation}
\label{Jcond}
\mathbf{J}|_{\partial\Gamma} = \mathbf{0} ,
\end{equation}
which follows from Eq.~\eqref{eq:7}.

The general governing dynamics for $\mathbf{R}(\chi,t)$ is given by:
\begin{equation}
\label{bdyng} \dot{\mathbf{R}} = \mathbf{U} ,
\end{equation}
where $\mathbf{U}(\chi,t)$ can be related to the contact line velocity as follows. Let $\bm{\Gamma}(\chi,t)$ denote the parametric form of the contact line (i.e., the 3-dimensional curve along which the liquid terminates). $\bm{\Gamma}(\chi,t)$ is related to the 2-dimensional domain boundary curve as:
\begin{equation}
\bm{\Gamma}(\chi,t) = (\mathbf{R}(\chi,t),\phi(\mathbf{R}(\chi,t))).   
\end{equation}
Note the distinction between $\bm{\Gamma}$ and $\partial\Gamma$; the first is the 3-dimensional contact line, while the second is its projection onto the $xy$-plane.
Therefore, the contact line velocity is
\begin{equation}
\label{eq:3D_vel}
\mathbf{V}(\chi,t)\equiv \dot{\bm{\Gamma}}(\chi,t) = (\mathbf{U},\nabla\phi|_{\partial\Gamma}\cdot\mathbf{U}),    
\end{equation}
while $\mathbf{U}(\chi,t)$ is the projection of the contact line velocity to the $xy$-plane.

To obtain expressions for $\mathbf{J}$ and $\mathbf{U}$, we invoke OVP \citep{onsager1931reciprocal_1, doi2011onsager, doi2013soft}, which is based on the Rayleighian of the system, defined in Eq.~\eqref{Rayleigh}.
Assuming constant temperature, the relevant thermodynamic potential of the system is the Helmholtz free energy---see e.g.~\cite{mandl1991statistical}.
At equilibrium the Helmholtz free energy $F$ is a minimum and the laws of thermodynamics imply that $F$ must always decrease in spontaneous non-equilibrium processes\footnote{We should emphasize that $F$ only decreases or remains constant over time in isolated systems. If the system is driven e.g.\ via a time dependent external potential or via the boundary conditions, then of course the free energy may increase over time.}.
The present OVP based approach naturally builds this physics into the description.

We assume that the time evolution of the liquid has no effect on the solid and the gas. This corresponds to the case when solid surface has a volume much larger than the volume of liquid upon it and it has a sizeable enough thermal conductivity that it acts as a heat reservoir, keeping the temperature in the whole system constant, and also that any motion in the air is slow enough that the forces it imposes on the liquid are negligible.
In this case, the Helmholtz free energy of the system can be written as a functional of the liquid thickness over the variable domain $\Gamma$:
\begin{equation}
\label{Fgen}F[h,\Gamma] = \int_{\Gamma} f(h,\nabla h)\,dA \,, 
\end{equation}
where we also assume that the free energy density $f(h,\nabla h)$ only depends on the liquid height and its gradient.
We say much more about the free energy below in \S~\ref{subsec:free_energy}.
Using the `transversality condition' extension of the calculus of variations for varying domain boundary \citep{gelfand1963calculus}, $\dot{F}$ can be written as (see Appendix~\ref{app:cv} for further details):
\begin{eqnarray}
\dot{F} &=& \int_{\Gamma} \left\{\dot{h}\,\frac{\delta F}{\delta h}\right\}\,dA + \int_0^1 \left\{ \dot{\mathbf{R}} \cdot \mathbb{T} \cdot \mathbf{n}_\perp \right\}d\chi\nonumber\\
&=& -\int_\Gamma \left\{(\nabla\cdot\mathbf{J}) \frac{\delta F}{\delta h} \right\} \,dA + \int_0^1 \left\{\mathbf{U}\cdot\,\mathbb{T}\cdot \mathbf{n}_\perp\right\} d\chi \,,\nonumber\\
&=& \int_\Gamma \left\{\mathbf{J} \cdot \nabla \frac{\delta F}{\delta h} \right\} \,dA + \int_0^1 \left\{\mathbf{U}\cdot\,\mathbb{T}\cdot \mathbf{n}_\perp\right\} d\chi \,, \label{dFdt}
\end{eqnarray}
where $\frac{\delta F}{\delta h}$ is the functional derivative of $F[h,\Gamma]$ with respect to $h(\mathbf{r},t)$, while
\begin{equation}
\label{BMA}
\mathbb{T} = \left( f\,\mathbb{I}-\nabla h \otimes \frac{\partial f}{\partial \nabla h} \right)_{\partial\Gamma},
\end{equation}
which we refer to as the `boundary matrix'.
In Eq.~\eqref{BMA} $\mathbb{I}$ is the $2 \times 2$ identity matrix and $\otimes$ denotes dyadic product.
To obtain Eq.~\eqref{BMA}, we have used the fact that $h=0$ at the contact line.

Returning to Eq.~\eqref{dFdt}, the first term on the right-hand side of the first line is the usual expression one obtains applying the chain rule for functionals, while the second term is due to the moving boundary at the contact line.
To obtain the second line, we use Eqs.~\eqref{hdyng} and \eqref{bdyng}, while the third line comes from an integration by parts and then using Eq.~\eqref{Jcond}, to note that the resulting boundary contribution is zero.

The dissipation functional $\Phi$ in Eq.~\eqref{Rayleigh} is not known exactly, but it must be positive definite on thermodynamic grounds \citep{onsager1931reciprocal_1}, and so the leading-order contributions must be quadratic in $\mathbf{J}$ and $\mathbf{V}$ (and so also $\mathbf{U}$).
In particular, we assume
\begin{equation}\label{eq:Phi}
    \Phi=\int_\Gamma \,\frac{|\mathbf{J}|^2}{2\,M_v}\,dA + \oint_{\partial\Gamma} \frac{|\mathbf{V}|^2}{2\,M_{\bm{\Gamma}}}\,dl,
\end{equation}
where $dl$ is the length of the contact line element (defined later)---not to be confused with the length of the domain boundary element. The first term in Eq.~\eqref{eq:Phi} is the viscous dissipation within the droplet, where $M_v>0$ is a mobility coefficient, while the second term is a dissipation (i.e.\ $M_{\bm{\Gamma}}>0$) associated with the motion of the contact line, which is assumed to be proportional to its length.
We defer saying any more about these two mobility coefficients till later.
In Appendix~\ref{app:df}, we present a derivation that illustrates the physics of Eq.~\eqref{eq:Phi}, which will be useful for readers with a background in fluid mechanics or those having less familiarity with thermodynamics.

Using the parametric form of the contact line, the dissipation functional can be written as:
\begin{equation}
\label{Phi}
\Phi = \int_\Gamma \,
\frac{|\mathbf{J}|^2}{2\,M_v}
\,dA
+ \int_0^1 \left(\frac{\mathbf{U} \cdot \mathbb{S} \cdot \mathbf{U}}{2\,M_{\bm{\Gamma}}}\,s\right)\,d\chi \,,
\end{equation}
where $s=\sqrt{\mathbf{R}_\chi\cdot\mathbb{S}\cdot \mathbf{R}_\chi}$ with $\mathbf{R}_\chi = (X_\chi(\chi,t),Y_\chi(\chi,t))$, while the matrix $\mathbb{S}=\mathbb{I}+(\nabla\phi \otimes \nabla\phi)|_{\partial\Gamma}$ encodes the geometry of the solid surface. The length of the contact line element corresponding to $d\chi$ is $dl = s\,d\chi$.

According to OVP, the system chooses a trajectory along which the Rayleighan \eqref{Rayleigh} is minimal with respect to variations in $\mathbf{J}$ and $\mathbf{U}$, i.e.\ the dynamical equations for the system are generated by the following pair:
\begin{equation}
\label{PLD}
\frac{\delta \mathcal{R}}{\delta\mathbf{J}} = 0 \quad \textrm{and} \quad \frac{\delta \mathcal{R}}{\delta\mathbf{U}} = 0 \,.
\end{equation} 
Using Eqs.~\eqref{Rayleigh}, (\ref{dFdt}) and (\ref{Phi}) in the first relation in Eq.~(\ref{PLD}) yields
\begin{equation}
\frac{\delta}{\delta \mathbf{J}}(\dot{F}+\Phi)=\nabla \frac{\delta F}{\delta h}+\frac{\mathbf{J}}{M_v}=0\,,
\end{equation}
which can be rearranged to give
\begin{equation}\label{eq:J}
    \mathbf{J} = -M_v\,\nabla \frac{\delta F}{\delta h}.
\end{equation}
Similarly, the second relation in Eq.~(\ref{PLD}) gives
\begin{equation}\label{eq:U}
\mathbf{U} = -M_{\bm{\Gamma}}\,\mathbb{S}^{-1}\cdot\mathbb{T}\cdot\mathbf{u}\,,
\end{equation}
where $\mathbf{u}(\chi,t)=\mathbf{n}_\perp/{s}$. Substituting these back into Eqs.~(\ref{hdyng}) and (\ref{bdyng}) results in the following general dynamical equations:
\begin{eqnarray}
\label{hdyn}
\dot{h} &=& \nabla \cdot\left( M_v\,\nabla \frac{\delta F}{\delta h} \right), \\
\label{bdyn}
\dot{\mathbf{R}} &=& -M_{\bm{\Gamma}}\,\mathbb{S}^{-1} \cdot \mathbb{T}\cdot\mathbf{u} \,.
\end{eqnarray}
The first is the variational form of the thin-film equation \citep{thiele2010thin, thiele2011note}, while the second is the equation for the time evolution of the contact line.
At this point it is worthwhile to note why we do not have a term $\sim \mathbf{J}\cdot\mathbf{U}$ in Eq.~\eqref{Phi}.
If present, this would give a term $\sim \mathbf{U}$ in Eq.~\eqref{eq:J}, which would couple the current $\mathbf{J}$ in the centre of the drop to the dynamics at the contact-line, which of course is incorrect for large droplets and films.
Moreover, if large enough, such a cross-term could render a quadratic $\Phi$ non-convex, breaking the positive-definite requirement $\Phi\geq0$, and also destroying the gradient dynamics structure.
That the OVP has resulted in a gradient flow is to be expected on thermodynamic grounds.
It is also what one expects if the underlying microscopic dynamics is time-reversible and Markovian \citep{mielke2014relation, MielkeRengerPeletier2016}.
Note also that one should not expect terms in $\Phi$ that are linear in any of the velocities, since this would correspond to a form of `active' driving \citep{wang2021onsager, archer2026variational}.

Inserting the results in Eqs.~(\ref{eq:J}) and (\ref{eq:U}) into Eq.~(\ref{dFdt}), it is straightforward to show that the free energy $F$ monotonically decreases or remains constant over time.
Doing this, we obtain
\begin{eqnarray}
\dot{F} &=& -\int_\Gamma \frac{\mathbf{J}^2}{M_v} \,dA - \int_0^1 \left(\frac{\mathbf{U} \cdot \mathbb{S} \cdot \mathbf{U}}{M_{\bm{\Gamma}}}\,s\right)\,d\chi \,. 
\label{dFdt_sub_in}
\end{eqnarray}
Since $M_v\geq0$, $M_{\bm{\Gamma}}\geq0$, $s>0$ and $\mathbb{S}$ is positive definite, from Eq.~\eqref{dFdt_sub_in} we see that $\dot{F} \leq 0$ for all $t$.

Due to volume conservation, we must also have $M_v|_{\partial\Gamma}=0$, to satisfy Eq.~(\ref{Jcond}).
Differentiating the terminal boundary condition $h(\mathbf{R}(\chi,t),t)=0$ with respect to $t$, then using Eqs.~(\ref{hdyn}) and (\ref{bdyn}) yields the following consistency criterion (dynamic boundary condition): 
\begin{equation}
\label{DBC}
\dot{h}|_{\partial\Gamma}+\nabla h|_{\partial\Gamma} \cdot \dot{\mathbf{R}} = 0 \,.
\end{equation} 
On the practical level, Eq.~\eqref{DBC} connects the spatial derivatives of $h(x,y,t)$ at $(x,y) \in \partial\Gamma(t)$.
For particular examples, see \S~\ref{sec:examples}. Whether Eq.~(\ref{DBC}) can be satisfied in general depends on the choice of $F$ and $M_v$.

We now discuss properties of the mobility coefficient $M_v$.
The mobility coefficient must be inversely proportional to the bulk liquid viscosity, $M_v\propto1/\eta$.
Additionally, in the limit $h\to0$, we should expect $M_v\to0$.
On dimensional grounds, we must have $M_v=\kappa{\cal L}^3/\eta$, where $\kappa$ is a dimensionless constant and ${\cal L}$ is a quantity that depends on $h$ and has the dimension of length.
Any further progress requires the physics that comes from taking the thin-film limit of the Navier Stokes equation \citep{oron1997long, craster2009dynamics}.
The liquid flow boundary condition with the solid surface determines the specific functional dependence of $M_v(h)$.
When there is a no-slip boundary condition, then one finds $M_v(h)=h^3/(3\eta)$.
In contrast, when there is slip and depending on the precise nature of the slip condition \citep{savva2011dynamics,haley1991effect}, we obtain $M_v(h)=h^{3-\nu}\,(h^\nu+3\lambda^\nu)/(3\eta)$, where $\lambda$ is the slip length and $\nu=1,2,3$.
The behaviour of Eq.~(\ref{DBC}) depends on the limiting behaviour of $h$ and its derivatives that enter into ${\delta F}/{\delta h}$ as $\partial\Gamma$ is approached. Any model where the $h\to0$ leading-order contribution to $M_v(h)$ is $h^q$, where $q \geq 2$, will either not move ($\dot{\mathbf{R}}=0$) or require resolution of singular derivatives of $h$ to obtain a finite $\dot{\mathbf{R}}$. Although the latter is possible numerically for Navier-slip where $q=2$ (for example, see \cite{savva2009two}), we choose the more amenable mobility function
\begin{equation}\label{eq:mobility}
M_v(h) \equiv h\,(h^2+\lambda^2) \,.
\end{equation}
Here we have dropped the bulk liquid viscosity factor $3\eta$, henceforth absorbing this into the timescale alongside an appropriate $\lambda$ scaling.
Note also that since $M_v$ incorporates the effect of the slip length $\lambda$, this builds in the effect of dissipation due to slip at both the contact line and also at the substrate; for further details see Appendix \ref{app:df}.
With this choice, we have $\nabla M_v|_{\partial\Gamma} = \lambda^2\,\nabla h|_{\partial\Gamma}$, which, when used in Eq.~(\ref{DBC}), yields
\begin{equation}
\label{DBCh}
    \nabla h|_{\partial\Gamma}\cdot\left\{ \lambda^2\,\left( \nabla \frac{\delta F}{\delta h}\right)\bigg|_{\partial\Gamma} + \dot{\mathbf{R}} \right\} = 0 \,.
\end{equation}
We should mention that because Eq.~\eqref{eq:mobility} is imposed in the thin-film limit, caution should be taken when applying the subsequent theory that follows to the case of drops or films on surfaces with large inclination angles.
In such cases, defining $h$ as a perpendicular distance and then using Eq.~\eqref{eq:mobility} may be more appropriate.

\subsection{Equilibrium}
\label{subsec:equilib}

The $t\to\infty$ stationary solutions of the dynamics occur when $\dot{h}=0$ and $\dot{\mathbf{R}}=\mathbf{0}$.
Equations \eqref{hdyn} and \eqref{bdyn} therefore yield
\begin{equation}
\label{EL} 
\frac{\delta F}{\delta h} = \press,
\end{equation}
and
\begin{equation}\label{eq:contact_line_equilib}
\mathbb{S}^{-1} \cdot \mathbb{T}\cdot\mathbf{u} = \mathbf{0}.    
\end{equation}
Equation (\ref{EL}) is the usual Euler-Lagrange equation for the equilibrium drop profile, where $\press$ is a (constant) Lagrange multiplier, which indicates that the stationary points of the dynamics coincide with those of the free energy at constant liquid volume.
As explained further below [see Eq.~\eqref{dFdh}], $\press$ is the equilibrium pressure in the liquid.
Equation \eqref{eq:contact_line_equilib} only has non-trivial solution for $\mathbf{u}$ if $\mathbb{T}$ is singular.
Since $\det(\mathbb{S}^{-1}) \neq 0$, this implies that at equilibrium we must have that
\begin{eqnarray}
\label{TRC} \det(\mathbb{T}) = 0,
\end{eqnarray}
which is generally known as the transversality condition.
We return to this below in \S~\ref{subsec:contact_angle} and show how this condition yields Young's equation.

\subsection{Free energy}
\label{subsec:free_energy}

To make Eqs.~(\ref{hdyn}) and (\ref{bdyn}) practical, we must have an explicit expression for the free energy $F$.
We derive here a simple approximation, incorporating gravity and capillarity.
If other physical phenomena are relevant, these should be incorporated via the free energy, extending the derivation we now present.
The free energy is made up of three contributions,
\begin{equation}
F = F_V + F_I + F_G \,,
\end{equation}
where $F_V$ is the bulk contribution depending on the volumes of the solid ($S$), liquid ($L$) and gas ($G$) in the system; $F_I$ is due to the presence of interfaces between each phase, and $F_G$ is associated with gravity.
Accordingly, $F_V = f_S V_S +f_L V_L + f_G V_G$, where $f_\alpha$ (constant) and $V_\alpha$ are the free energy density and volume of phase $\alpha=S,L,G$.
The interfacial contribution $F_I = \gamma_{LG}\,A_{LG} + \gamma_{SL}\,A_{SL} + \gamma_{SG}\,A_{SG}$, where $\gamma_{\alpha\beta}$ (constant) are the interfacial tensions (interfacial excess free energies) between phases $\alpha$ and $\beta$ and $A_{\alpha\beta}$ are the areas of these interfaces.
The gravitational contribution is $F_G=\int\rho(x,y,z,t)gz dV$, where $g$ is the gravitational acceleration and $\rho$ is the density distribution given in Eq.~\eqref{eq:rho_cases}. In the absence of evaporation, condensation or any other phase transitions, the three volumes $V_\alpha$ are constant, with the total volume $V = V_S + V_L + V_G$ also constant.
Consequently, $F_V=C_1$ (constant).
As the solid is stationary, $A_{SL}+A_{SG}$ is also constant, and therefore adding and subtracting $\gamma_{SG}(A_{SG}+A_{SL})$ (constant) to/from $F_I$ results in $F_I = \gamma_{LG}\,A_{LG} + (\gamma_{SL}-\gamma_{SG})\,A_{SL} + C_2$, where $C_2$ is constant. Finally, since only the liquid and gas phases are moving, we can write $F_G = \int_{V_L} \Delta\rho_0\,gz\,dV + C_3$ where $C_3$ is constant and $\int_{V_L}$ denotes an integral over the volume containing the liquid (recall $\Delta\rho_0=\rho_L-\rho_G$).
Collecting everything together yields the excess free energy
\begin{eqnarray}\label{eq:Delta_F}
\Delta F &\equiv& F - F_0 \\
&=& \gamma_{LG}\,A_{LG} + (\gamma_{SL}-\gamma_{SG})\,A_{SL} + \Delta\rho_0\,g \int_{V_L} z\,dV \,, \nonumber
\end{eqnarray} 
where $F_0=C_1+C_2+C_3$. The area of the solid-liquid interface can be written as
\begin{equation}
A_{SL} = \int_\Gamma \sqrt{1+|\nabla\phi|^2}\,dA,    
\end{equation}
while that of the liquid-gas interface is
\begin{equation}
A_{LG} = \int_\Gamma \sqrt{1+|\nabla \xi|^2}\,dA,
\end{equation}
where $\xi(x,y,t)$, the position of the liquid-gas interface, is defined in Eq.~\eqref{eq:xi}. Furthermore, the gravitational contribution
\begin{eqnarray}
\Delta F_G &=& \Delta\rho_0g\,\int_\Gamma \int_\phi^{\xi} z\,dz\,dA \nonumber \\
&=& \Delta\rho_0g \int_\Gamma \,\frac{\xi^2-\phi^2}{2}\,dA.
\end{eqnarray}
A dimensionless form of the free energy can be obtained by introducing the length scale $\ell$ and energy scale $\gamma_{LG}\,\ell^2$, yielding:
\begin{equation}
\label{Fnond} 
F[h,\Gamma] =  \int_{\Gamma} \left\{ \sqrt{1+|\nabla \xi|^2} - \PP\,\sqrt{1+|\nabla\phi|^2} + \frac{B}{2}\,(\xi^2-\phi^2)\right\}\,dA\,,
\end{equation}
where now $F$ denotes the dimensionless excess free energy $F[h,\Gamma] = \Delta F/(\gamma_{LG}\,\ell^2)$.
Likewise, $\xi(x,y,t)$ and $\phi(x,y)$ are dimensionless lengths, measured in $\ell$ units\footnote{Note that we thus retain the full curvature expressions for the interfacial tension contributions, rather than making the usual thin-film equation `long-wave' approximation $\sqrt{1+|\nabla h|^2}\approx 1+(1/2)|\nabla h|^2$, while at the same time using Eq.~\eqref{eq:mobility}, which is typically imposed in the long-wave approximation \citep{oron1997long}. Such combinations have been important in other thin-film problems -- see e.g.\ \cite{wilson1982drag, snoeijer2006free, Thie2018csa, lopes2018multiple, diekmann2025mesoscopic}.}.
The remaining dimensionless model parameters are the Bond number $B = \Delta\rho_0 g\ell^2/\gamma_{LG}$ and $\PP=(\gamma_{SG}-\gamma_{SL})/\gamma_{LG}$.
Note that the term in Eq.~\eqref{Fnond} with prefactor $\PP$ is needed because $\Gamma$ varies over time.
The equations for the dynamics, Eqs.~(\ref{hdyn}), (\ref{bdyn}) and (\ref{DBCh}), require the functional derivative
\begin{equation}
\label{dFdh} \frac{\delta F}{\delta h} = B\,\xi - \nabla\cdot \left(\frac{\nabla \xi}{\sqrt{1+|\nabla \xi|^2}}\right)
\end{equation}
and the boundary matrix \eqref{BMA} becomes
\begin{eqnarray}
\label{Bmat}
\mathbb{T} = \Bigg[\left( \sqrt{1+|\nabla \xi|^2}-\PP\,\sqrt{1+|\nabla\phi|^2} \right) \,\mathbb{I}
- \frac{\nabla h \otimes \nabla \xi}{\sqrt{1+|\nabla \xi|^2}} \Bigg]_{\partial\Gamma} \,.
\end{eqnarray}
Note that because the boundary matrix is evaluated at $\partial\Gamma$ where $h=0$ and $\xi=\phi$, the term with coefficient $B$ is also zero.
Providing a physical interpretation of Eq.~(\ref{Bmat}) requires further analysis.
Note also that the first term in Eq.~\eqref{dFdh} is the hydrostatic pressure, while the second term is the Laplace pressure [c.f.~Eq.~\eqref{EL}].

\subsection{Contact angle}
\label{subsec:contact_angle}

The projected contact line dynamics given by Eq.~(\ref{bdyn}) can be understood by expressing the boundary matrix \eqref{Bmat} in terms of the dynamic contact angle $\theta (\chi,t)$.
We now start from Eqs.~(\ref{bdyn}) and \eqref{Bmat} and follow a derivation resulting in showing that at equilibrium Young's equation is satisfied and that away from equilibrium, the contact line speed $|\mathbf{V}|\propto(\cos\theta-\cos\theta_s)$, where $\theta_s$ is the equilibrium contact angle.
Some readers may want to skip forward to the final results in Eqs.~\eqref{Young} and \eqref{eq:V_magnitude_2}, while others will want all the intervening steps.

By definition, at a given point along the contact line, the dynamic contact angle $\theta(\chi,t)$ is the dihedral angle between the tangent planes to the surfaces $z=\xi(x,y,t)$ and $z=\phi(x,y)$ at that point.
Consequently, $\theta(\chi,t)$ satisfies the equation
\begin{equation}
\label{cangle}
\cos\theta(\chi,t) = \frac{\mathbf{n}_\xi \cdot \mathbf{n}_\phi}{|\mathbf{n}_\xi|\,|\mathbf{n}_\phi|} \,,
\end{equation}
where $\mathbf{n}_\xi \equiv (-\xi_x,-\xi_y,1)|_{\partial\Gamma}$ and $\mathbf{n}_\phi \equiv (-\phi_x,-\phi_y,1)|_{\partial\Gamma}$ are the normals to the surfaces $z=\xi(x,y,t)$ and $z=\phi(x,y)$, respectively, at the given point on the contact line.
Using Eq.~(\ref{cangle}) together with $|\mathbf{n}_\xi|=\sqrt{1+|\nabla \xi|^2}$ and $|\mathbf{n}_\phi|=\sqrt{1+|\nabla \phi|^2}$, we can consider in turn the four components of the matrix $\mathbb{T}$ in Eq.~\eqref{Bmat}.
These are:
\begin{eqnarray}
    \mathbb{T}_{11}&=&|\mathbf{n}_\xi|-\PP|\mathbf{n}_\phi|-\frac{h_x\xi_x}{|\mathbf{n}_\xi|} \nonumber\\
    &=&\frac{|\mathbf{n}_\xi|^2-(\xi_x-\phi_x)\xi_x}{|\mathbf{n}_\xi|}-\PP|\mathbf{n}_\phi|\nonumber\\
    &=&\frac{1+\xi_x^2+\xi_y^2-\xi_x^2+\phi_x\xi_x}{|\mathbf{n}_\xi|}-\PP|\mathbf{n}_\phi|\nonumber\\
    &=&\frac{1+\phi_x\xi_x+(\phi_y\xi_y-\phi_y\xi_y)+\xi_y^2}{|\mathbf{n}_\xi||\mathbf{n}_\phi|}|\mathbf{n}_\phi|-\PP|\mathbf{n}_\phi|\nonumber\\
    &=&\left(\frac{\mathbf{n}_\phi\cdot\mathbf{n}_\xi+h_y\xi_y}{|\mathbf{n}_\xi||\mathbf{n}_\phi|}-\PP\right)|\mathbf{n}_\phi|\nonumber\\
    &=&\left(\cos\theta-\PP\right)|\mathbf{n}_\phi|+\frac{h_y\xi_y}{|\mathbf{n}_\xi|}
    \label{eq:T_11}
\end{eqnarray}
where we used Eq.~\eqref{eq:xi} in obtaining the second and fifth lines.
The two off-diagonal terms of \eqref{Bmat} are
\begin{eqnarray}
    \mathbb{T}_{12}=-\frac{h_x\xi_y}{\sqrt{1+|\nabla\xi|^2}}
    =-\frac{h_x\xi_y}{|\mathbf{n}_\xi|}
    \label{eq:T_12}
\end{eqnarray}
and
\begin{eqnarray}
    \mathbb{T}_{21}=-\frac{h_y\xi_x}{\sqrt{1+|\nabla\xi|^2}}
    =-\frac{h_y\xi_x}{|\mathbf{n}_\xi|}.
    \label{eq:T_21}
\end{eqnarray}
Finally, following a derivation exactly analogous to that in Eq.~\eqref{eq:T_11}, the remaining component is
\begin{eqnarray}
    \mathbb{T}_{22}=\left(\cos\theta-\PP\right)|\mathbf{n}_\phi|+\frac{h_x\xi_x}{|\mathbf{n}_\xi|}.
    \label{eq:T_22}
\end{eqnarray}
Combining Eqs.~\eqref{eq:T_11}--\eqref{eq:T_22} into a single matrix leads to
\begin{equation}
\label{Tmat}
\mathbb{T} = |\mathbf{n}_\phi|(\cos\theta-\PP)\,\mathbb{I} + \frac{1}{|\mathbf{n}_\xi|}\,\left.\left( \begin{matrix} \xi_y\,h_y & -\xi_y\,h_x \\ -\xi_x\,h_y & \xi_x\,h_x \end{matrix} \right)\right|_{\partial\Gamma} \,.
\end{equation}
With this expression for $\mathbb{T}$, we now return to consider the contact line velocity $\mathbf{V}$.
Squaring the expression for this in Eq.~\eqref{eq:3D_vel} gives
\begin{eqnarray}
|\mathbf{V}|^2&=&\dot{\mathbf{R}}^T\cdot\dot{\mathbf{R}}+(\nabla\phi|_{\partial\Gamma}\cdot\dot{\mathbf{R}})^2\nonumber\\
&=&\dot{\mathbf{R}}^T\cdot\dot{\mathbf{R}}+\dot{\mathbf{R}}^T(\nabla\phi|_{\partial\Gamma}\otimes\nabla\phi|_{\partial\Gamma})\cdot\dot{\mathbf{R}}\nonumber\\
&=&\dot{\mathbf{R}}^T\cdot(\mathbb{I}+\nabla\phi|_{\partial\Gamma}\otimes\nabla\phi|_{\partial\Gamma})\cdot\dot{\mathbf{R}}\nonumber\\
&=&\dot{\mathbf{R}}^T\cdot\mathbb{S}\cdot\dot{\mathbf{R}},
\label{eq:V_squared}
\end{eqnarray}
where the superscript $T$ denotes the transpose.
Previously, we have neglected to differentiate between a vector and its transpose, for ease of notation.
However, at this point of our discussion it is worth indicating the difference.
Inserting Eq.~\eqref{bdyn} into Eq.~\eqref{eq:V_squared} gives
\begin{eqnarray}
   |\mathbf{V}|^2&=&M_{\bm{\Gamma}}^2(\mathbb{S}^{-1}\cdot\mathbb{T}\cdot\mathbf{u})^T\cdot\mathbb{S}\cdot(\mathbb{S}^{-1}\cdot\mathbb{T}\cdot\mathbf{u})\nonumber \\
   &=&M_{\bm{\Gamma}}^2(\mathbb{T}\cdot\mathbf{u})^T\cdot(\mathbb{S}^{-1})^T\cdot(\mathbb{S}\cdot\mathbb{S}^{-1})\cdot(\mathbb{T}\cdot\mathbf{u})\nonumber \\
   &=&M_{\bm{\Gamma}}^2(\mathbb{T}\cdot\mathbf{u})^T\cdot\mathbb{S}^{-1}\cdot(\mathbb{T}\cdot\mathbf{u}),
   \label{eq:V_squared_2}
\end{eqnarray}
where we have used the fact that the matrix $\mathbb{S}$ is symmetric.

Since the domain boundary curve $\partial\Gamma$ is a contour line of $h$ (due to the terminal boundary condition $h|_{\partial\Gamma}=0$), we have that $\mathbf{u}\propto \nabla h$ at any point along $\partial\Gamma$, allowing us to write $\mathbf{u}=(\frac{|\mathbf{n}_\perp|}{s|\nabla h|})\nabla h|_{\partial\Gamma}$ -- see Eq.~\eqref{PLD}.
So, with Eq.~\eqref{Tmat} we obtain
\begin{equation}\label{eq:Tdotu}
    \mathbb{T}\cdot\mathbf{u}=\left(\frac{|\mathbf{n}_\perp|}{s|\nabla h|}\right)|\mathbf{n}_\phi|(\cos\theta-\PP)\nabla h|_{\partial\Gamma}. 
\end{equation}
Substituting this into Eq.~\eqref{eq:V_squared_2} gives
\begin{eqnarray}
   |\mathbf{V}|^2&=&\left[M_{\bm{\Gamma}}\left(\frac{|\mathbf{n}_\perp|}{s|\nabla h|}\right)|\mathbf{n}_\phi|(\cos\theta-\PP)\right]^2(\nabla h\cdot\mathbb{S}^{-1}\cdot\nabla h)|_{\partial\Gamma}\nonumber \\
   &=&\left[\frac{M_{\bm{\Gamma}}|\mathbf{n}_\perp|}{s}(\cos\theta-\PP)\right]^2\left[\frac{\nabla h}{|\nabla h|}\cdot(|\mathbf{n}_\phi|^2\mathbb{S}^{-1})\cdot\frac{\nabla h}{|\nabla h|}\right]\nonumber \\
   &=&\left[\frac{M_{\bm{\Gamma}}}{s}(\cos\theta-\PP)\right]^2\left(\mathbf{n}_\perp\cdot(|\mathbf{n}_\phi|^2\mathbb{S}^{-1})\cdot\mathbf{n}_\perp\right),
   \label{eq:V_squared_next}
\end{eqnarray}
where we have used the fact that $\nabla h/|\nabla h|=\mathbf{n}_\perp/|\mathbf{n}_\perp|$ along $\partial\Gamma$.
Note also that
\begin{eqnarray}
    |\mathbf{n}_\phi|^2\mathbb{S}^{-1}&=&(1+\phi_x^2+\phi_y^2)|_{\partial\Gamma}\mathbb{S}^{-1}\nonumber\\
    &=&\mathbb{I}+\left.
\begin{pmatrix}
\phi_y^2 & -\phi_x\phi_y \\
-\phi_x\phi_y & \phi_x^2 
\end{pmatrix}\right|_{\partial\Gamma}
\end{eqnarray}
and so $\mathbf{n}_\perp\cdot(|\mathbf{n}_\phi|^2\mathbb{S}^{-1})\cdot\mathbf{n}_\perp=\mathbf{n}_\parallel\cdot\mathbb{S}\cdot\mathbf{n}_\parallel=s^2$, finally giving
\begin{equation}\label{eq:V_magnitude}
    |\mathbf{V}|=M_{\bm{\Gamma}} |\cos\theta-\PP|.
\end{equation}
Therefore, the equilibrium condition (\ref{TRC}), which is equivalent to $|\mathbf{V}|=0$, equates to having $\cos\theta=\PP$ at all points along the contact line.
Recalling that $\PP=(\gamma_{SG}-\gamma_{SL})/\gamma_{LG}$ (see above \eqref{dFdh}), we now can see that this is identical to having Young's equation
\begin{equation}
\label{Young} \cos\theta_s = \frac{\gamma_{SG}-\gamma_{SL}}{\gamma_{LG}} \,,
\end{equation}
for all points along the contact line, where $\theta_s$ is the static equilibrium value of $\theta(\chi,t)$.
Since $\PP=\cos\theta_s$, we may rewrite Eq.~\eqref{eq:V_magnitude} as
\begin{equation}\label{eq:V_magnitude_2}
    |\mathbf{V}|=M_{\bm{\Gamma}} |\cos\theta-\cos\theta_s|.
\end{equation}
This result is physically reassuring, because it does not include any factors related to the difference between the physical contact line $\bm{\Gamma}$ and its projection onto the $xy$-plane, $\partial\Gamma$.
It also adds to the growing consensus \citep{RenE,andreas2024} that a significant source of dissipation in contact line motion arises from the deviation of the dynamic contact angle from its equilibrium value as $(\cos\theta-\cos\theta_s)$ in theories such as Molecular Kinetic Theory (MKT) \citep{BLAKE1969421}, experiments \citep{Duvivier,Ramiasa} and Dynamic Density Functional Theory (DDFT) \citep{andreas2024}. Numerous authors have used either this condition or a similar velocity dependent one, such as in \cite{greenspan1978motion, haley1991effect, hocking1992rival, wilson2000rate, sibley2015comparison}, and whilst there has been debate, it is of significant note to see this result appearing naturally from the minimisation of the Rayleighan alongside the thin-film equation in a self-consistent fashion.

Like the bulk mobility coefficient $M_v$, we should expect the contact line mobility coefficient $M_{\bm{\Gamma}}$ to be inversely proportional to the bulk viscosity \citep{guo2013direct}, remembering that viscosity has been absorbed into the timescale and likewise the surface tension into the energy scale.
Since contact line motion is driven by capillary forces, we also expect $M_{\bm{\Gamma}}$ to be proportional to the liquid-gas surface tension $\gamma_{LG}$.
On dimensional grounds, one should also expect $M_{\bm{\Gamma}}\propto\gamma_{LG}/\eta$, since this quantity has the dimensions of velocity.
We should also expect $M_{\bm{\Gamma}}$ to depend on the local contact angle.
Following de Gennes' argument \citep{de1986deposition} leads to $M_{\bm{\Gamma}}\propto\theta$, and therefore here we use
\begin{equation}\label{eq:contact_line_mob}
    M_{\bm{\Gamma}}=M_{\bm{\Gamma}}^0\,\theta,
\end{equation}
where $M_{\bm{\Gamma}}^0\propto\gamma_{LG}/\eta$ is a constant\footnote{There are other possibilities for how $M_{\bm{\Gamma}}$ depends on $\theta$. For example, following the arguments in the Appendix of \cite{sarlin2025macroscopic} leads instead to $M_{\bm{\Gamma}}=M_{\bm{\Gamma}}^0\,\tan\theta$.}.
As Eq.~\eqref{eq:V_magnitude_2} (and also Eq.~\eqref{eq:contact_line_mob}) applies to the microscopic (actual) contact angle at the substrate, it remains consistent with well-known results for spreading droplets such as the behaviour of the macroscopic/apparent angle in the Cox-Voinov law \citep{Cox_1986,voinov1976hydrodynamics,snoeijer2013moving}, and spreading from Tanner's law \citep{tanner1979spreading}---see \S~\ref{sec:liq_ridge} below.

\section{Including evaporation}
\label{sec:evap}

We now extend the previous arguments to the situation when there is evaporation/condensation of the liquid to/from the gas phase.
In this case, the continuity equation \eqref{hdyng} becomes
\begin{equation}\label{eq:continuity_nc}
\dot{h}=-\nabla\cdot\mathbf{J}+\sigma,
\end{equation}
where $\sigma(\mathbf{r},t)$ is the evaporation/condensation flux at the surface of the liquid at point $\mathbf{r}$ at time $t$.
When $\sigma$ is negative, this corresponds to evaporation, and when positive, corresponds to condensation.
Again, we assume the system is isothermal, i.e.\ the solid surface is acting as a heat source/sink, keeping the temperature in the liquid and gas fixed.
This common assumption \citep{oron1997long}, is relevant whenever heat conduction through the thin film is faster than the fluid flow processes, so any latent heat generated by evaporation is quickly absorbed by the substrate.
The relevant thermodynamic potential is now the (dimensionless) grand free energy
\begin{equation}\label{eq:Omega}
    \Omega = F-p_G\int_\Gamma h \,dA\,,
\end{equation}
where $F$ is the Helmholtz free energy \eqref{Fnond} and the constant $p_G=\mu_G\rho_L/(m\gamma_{LG}\ell^2)$, where $\mu_G$ is the chemical potential of the vapour (gas) phase (strictly speaking, $\mu_G$ is the chemical potential difference between the value in the vapour and the value at gas-liquid phase coexistence \citep{dietrich88}) and $m$ is the mass of a liquid molecule, so $\rho_L/m$ is the number density of the liquid phase.
We are now treating the gas phase as being a (sufficiently large) reservoir, supplying particles to/from the liquid, without this leading to a change in the chemical potential of the gas.
The second term in Eq.~\eqref{eq:Omega} equals $-\mu_GN_L/(\gamma_{LG}\ell^2)$, where $N_L$ is the total number of molecules in the liquid.
The constant $p_G$ may be thought of as the Lagrange multiplier determining the $t\to\infty$ volume of liquid on the surface, when it equilibrates with the vapour phase with chemical potential $\mu_G$.
Owing to the way $p_G$ is related to $\mu_G$, it is also the pressure difference between that of the surrounding vapour phase and the value at gas-liquid phase coexistence. Thus, as we see in the following, for a flat film the sign of $p_G$ determines whether liquid will either evaporate into the vapour phase (negative $p_G$), or will condense from the vapour onto the surface (positive $p_G$).

The dissipation functional is now (c.f.~Eqs.~\eqref{eq:Phi}--\eqref{Phi} and note the analogous quadratic structure)
\begin{equation}
\Phi=\int \frac{|\mathbf{J}|^2}{2M_v}dA
+ \int_0^1 \left(\frac{\mathbf{U} \cdot \mathbb{S} \cdot \mathbf{U}}{2\,M_{\bm{\Gamma}}}\,s\right)\,d\chi \,
+\int \frac{\sigma^2}{2M_{nc}}dA\,,
\label{eq:Phi_grand}
\end{equation}
where $M_{nc}\geq0$ denotes the non-conserved (evaporative) dynamics mobility coefficient \citep{MoHo1980jcis, oron1997long, Ajae2005pre, thiele2009modelling, thiele2010thin}.
Note that we are developing here a `one sided' model, assuming that any fluxes within the gas are negligible.
Including vapour diffusive fluxes would lead to the theory of \cite{hartmann2023sessile}, which (owing to its gradient dynamics structure) may also be recast as a theory generated by OVP.
However, we refrain from pursuing that connection here. There is a substantial body of literature on evaporating droplets including the excellent review of \cite{HannahStephenEvapReview}. It is also possible, as for the other mobility coefficients, that there could be a dependence on $h$, $\theta$, etc., such as for the (somewhat different) models considered in \cite{Colinet,Savva_Rednikov_Colinet_2017} amongst others. For simplicity, in the example considered below in \S~\ref{subsec:evap_example}, we use a constant value.

The Rayleighan \eqref{Rayleigh} becomes ${\cal R}=\dot{\Omega}+\Phi$ and OVP dictates that the system chooses a trajectory along which ${\cal R}$ is minimal with respect to variations in $\mathbf{J}$, $\mathbf{U}$ and $\sigma$ [c.f.~Eq.~\eqref{PLD}], giving:
\begin{equation}
\label{PLD_evap}
\frac{\delta \mathcal{R}}{\delta\mathbf{J}} = 0, \quad\frac{\delta \mathcal{R}}{\delta\mathbf{U}} = 0 \quad \textrm{and} \quad \frac{\delta \mathcal{R}}{\delta\sigma} = 0 \,.
\end{equation}
The time derivative of the free energy is
\begin{align}\nonumber
    \dot{\Omega}&=\int\frac{\delta\Omega}{\delta h}\dot{h} \,dA+ \int_0^1 \left\{\mathbf{U}\cdot\,\mathbb{T}\cdot \mathbf{n}_\perp\right\} d\chi \,\\
&=\int\frac{\delta\Omega}{\delta h}(-\nabla\cdot\mathbf{J}+\sigma)\, dA+ \int_0^1 \left\{\mathbf{U}\cdot\,\mathbb{T}\cdot \mathbf{n}_\perp\right\} d\chi \,\nonumber\\&=\int\left(\nabla\frac{\delta\Omega}{\delta h}\cdot\mathbf{J}+\frac{\delta\Omega}{\delta h}\sigma\right)\, dA+ \int_0^1 \left\{\mathbf{U}\cdot\,\mathbb{T}\cdot \mathbf{n}_\perp\right\} d\chi \,\,.
\label{eq:Omega_dot}
\end{align}
The first line above is simply the chain rule for functionals with the second term being the transversality condition (entirely analogous to Eq.~\eqref{dFdt}).
The second line comes from using Eq.~\eqref{eq:continuity_nc} and the third comes from integrating by parts, assuming the boundary contributions are zero (i.e.\ assuming Eq.~\eqref{Jcond} applies).
Using Eqs.~\eqref{eq:Phi_grand} and \eqref{eq:Omega_dot}, together with the first condition in Eq.~\eqref{PLD_evap} yields
\begin{equation}
\frac{\delta}{\delta \mathbf{J}}(\dot{\Omega}+\Phi)=\nabla \frac{\delta \Omega}{\delta h}+\frac{\mathbf{J}}{M_v}=0\,,
\end{equation}
which gives [c.f.~Eq.~\eqref{eq:J}]
\begin{equation}\label{eq:J_nc}
\mathbf{J}=-M_v\nabla \frac{\delta \Omega}{\delta h}\,.
\end{equation}
The second condition in Eq.~\eqref{PLD_evap} gives again the contact line dynamical equation \eqref{bdyn}, with the modified boundary matrix [c.f.~Eq.~\eqref{BMA}]
\begin{equation}
\label{BMA_grand}
\mathbb{T} = \left( \omega\,\mathbb{I}-\nabla h \otimes \frac{\partial \omega}{\partial \nabla h} \right)_{\partial\Gamma},
\end{equation}
where the grand potential density $\omega=f-p_Gh$, while the third condition in Eq.~\eqref{PLD_evap} gives
\begin{equation}\label{eq:sigma}
\sigma=-M_{nc}\frac{\delta \Omega}{\delta h}\,,
\end{equation}
which can be re-written as $\sigma=-M_{nc}(p_L-p_G)$, where $p_L$ and $p_G$ are the pressure differences in the liquid and vapour, respectively -- see Eqs.~\eqref{eq:Omega} and \eqref{EL}. 
Substituting Eqs.~\eqref{eq:J_nc} and \eqref{eq:sigma} into Eq.~\eqref{eq:continuity_nc}, we obtain
\begin{equation}\label{eq:evap_dyn}
\dot{h}=\nabla\cdot\left[M_v\nabla \frac{\delta \Omega}{\delta h}\right]-M_{nc}\frac{\delta \Omega}{\delta h}\,.
\end{equation}
This is the slow evaporation rate (isothermal) dynamical equation used previously with much success to understand liquid film evaporation dynamics in various different contexts \citep{samid1998pattern, padmakar1999instability, lyushnin2002fingering, thiele2009modelling, thiele2010thin, frastia2011dynamical, frastia2012modelling}.
Note also that in the limit when $M_{nc}=0$, this reduces to the usual variational formulation of the thin-film equation.

Finally, as for the case without evaporation, 
Eq.~\eqref{DBC} holds as a consistency condition.
However, on using \eqref{eq:evap_dyn}, we obtain the following modified dynamic boundary condition 
\begin{equation}
\label{evapDBCh}
    \nabla h|_{\partial\Gamma}\cdot\left\{ \lambda^2\,\left( \nabla \frac{\delta \Omega}{\delta h}\right)\bigg|_{\partial\Gamma} + \dot{\mathbf{R}} \right\} = M_{nc}
    \frac{\delta \Omega}{\delta h}\bigg|_{\partial\Gamma} \,.
\end{equation}

\section{Some simple cases}
\label{sec:examples}

In this section, we discuss various limiting cases of the above general approach, in order to illustrate what the theory can do and to connect to known results in the literature and thereby validate our approach.

\subsection{Small liquid ridge on a flat homogenous surface}
\label{sec:liq_ridge}

Arguably the simplest prototype example of contact line motion is that of a small liquid ridge spreading on a homogenous surface (often referred to as a two-dimensional sessile droplet), where the cross-sectional area is small enough to be able to neglect gravity.
This occurs when the drop height is much smaller than the capillary length, so we have $B=0$.
This and closely related situations have been extensively studied \citep{hocking1992rival,greenspan1978motion,savva2009two,savva2011dynamics,haley1991effect,Hocking83}, and constitute an ideal test case of the OVP-based approach and for the numerical scheme used to solve the resulting equations, which is described in Appendix~\ref{app:num_ridge}.

The ridge height varies in the $x$-direction only---reducing to $h(x,t)$, and the flat and horizontal substrate means that $\phi=0$ (and $h(x,t)$=$\xi(x,t)$). Neglecting gravity, the free energy is simply
\begin{equation}
\label{eq:eg1fullF}
\frac{F}{L} = \int_\Gamma \left\{ \sqrt{1+h_x^2} - \PP \right\} \,dx \,,
\end{equation}
where $L$ is the length of the liquid ridge in the perpendicular $y$-direction. It is important to retain the (constant) $\PP$ here even though only derivatives of $F$ enter the governing equations since the (wet) contact area changes in time and hence the integral of $\PP$ plays a role in $\dot{F}$. The contact area $\Gamma(t)$ is a strip with two moving parallel sides of length $L$ characterised by $\mathbf{R}_\pm(\chi,t) = (X_\pm(t),Y(\chi))$, where $X_-(t) \leq X_+(t)$.
Recall that the surface tension has already been scaled out in the general case in Eq.~\eqref{Fnond} so enters in $\PP$ instead of as a prefactor of the interfacial term.
One further simplification we now exploit in order to obtain an asymptotic solution is to consider the long-wave
approximation where small-slopes $|h_x| \ll 1$ mean the free energy can be approximated as
\begin{equation}
\label{eq:eg1LW}
\frac{F_{\textrm{LW}}}{L} = \int_\Gamma \left\{\left(1+\frac{1}{2}h_x^2\right) -\PP \right\}\,dx \,.
\end{equation}
The contact lines at $x=X_\pm$ and corresponding contact angles are given by $\mp \tan(\theta_\pm) = h_x|_{x=R_\pm}$, which in the long-wave approximation simplifies to $\mp \theta_{\pm} = h_x|_{x=R_\pm}$.

Using the full curvature free energy from Eq.~\eqref{eq:eg1fullF}, the evolution equations (\ref{hdyn}) and (\ref{bdyn}) with (\ref{eq:mobility}) become
\begin{align}
\label{hasympsqrt} \dot{h} &= \frac{\partial}{\partial x} \left[ -h\,(h^2+\lambda^2) \frac{\partial^2}{\partial x^2} \left( \frac{h_x}{\sqrt{1+h_x^2}} \right) \right] , \\
\label{basympsqrt} \dot{X}_{\pm} &= \mp\,M_{\bm{\Gamma}}^0 \,\theta_{\pm}[\cos\theta_{\pm}-\cos\theta_{\mathrm{s}}] \,,
\end{align}
or in the long-wave approximation using Eq.~\eqref{eq:eg1LW} we have
\begin{align}
\label{hasympsqrtLW} \dot{h} &= \frac{\partial}{\partial x} \left[ -h\,(h^2+\lambda^2) \frac{\partial^3 h}{\partial x^3} \right] , \\
\label{basympsqrtLW} \dot{X}_{\pm} &= \pm\,\frac{M_{\bm{\Gamma}}^0 \,\theta_{\pm}}{2} [\theta_{\pm}^2-\theta_{\mathrm{s}}^2] \,,
\end{align}
where the $\mp$ cases on the right-hand side in Eq.~(\ref{basympsqrt}) are due to the fact that $\mathbf{u}=(1,0)$ on $x=X_+(t)$ and $\mathbf{u}=(-1,0)$ on $x=X_-(t)$.
Note also the sign change in Eq.~(\ref{basympsqrtLW}) since $\cos\theta=1-\theta^2/2+\ldots$ in the long-wave regime. Finally, the dynamic boundary condition can be obtained by imposing Eq.~\eqref{eq:eg1fullF} and the chosen geometry here on Eq.~(\ref{DBCh}) (see Appendix~\ref{app:dbc} for details), thus yielding
\begin{equation}
\label{eq:eg1dbc}
    \frac{\dot{X}_{\pm}}{\lambda^2} = \cos^3\theta_\pm\, \left[ h_{xxx} \pm\,\frac{3}{2}\,\sin(2\,\theta_\pm)\,h_{xx}^2 \right],
\end{equation}
where the derivatives of $h$ are evaluated at $x=X_\pm(t)$, and we do not require an equivalent in the long-wave form for the asymptotic procedure (where it is simply $\dot{X}_{\pm} = \lambda^2 h_{xxx}|_{x=X_\pm}$).

The asymptotic setup to employ for the long-wave version of the model is a singular perturbation problem in $\lambda$, where an outer region away from the contact lines has $\lambda=0$ but inner regions of $O(\lambda)$ near $x=X_\pm$ are required to satisfy the contact angle boundary condition.
Within this setup, a regular perturbation for slow (quasi-static) motion ($|\dot{X}_\pm|\ll1$) is exploited.
This approach has been extensively used \citep{Hocking83,savva2009two,sibley2015comparison,sibley2015asymptotics}, so is only briefly described here.

In the outer region, the leading-order (static) profile is found from the solution of $h_{xxx}=0$, i.e.~a parabolic profile, subject to $h|_{x=X_\pm}=0$ for touchdown and the cross-sectional area constraint $A = \int_{x=X_-}^{x=X^+} h\,dx$, which leads to
\begin{equation}
\label{eq:parabola}
   h = \frac{6A}{(X_+ - X_-)^3} (X_+ - x)(x-X_-).
\end{equation}
We consider the case of a symmetric drop (which occurs in this setup provided the initial condition is symmetric), and without further loss of generality, we can shift the ridge location so that the centre is located on $x=0$ with contact lines at $x=\pm a$ instead of $x=(X_+ + X_-)/2$.
Thus, in this form the profile is given by
\begin{equation}
   h = \frac{3A}{4a^3}(a^2-x^2).
\end{equation}
Remaining in the outer region, the next order in the quasi-static expansion necessitates a correction to $h$ at $O(\dot{a})$. Following \cite{sibley2015comparison} closely (the only substantive difference in the outer region being that we leave the cross-sectional area of the ridge $A$ unspecified rather than having it scaled out) we find 
\begin{align}
   h = \frac{3A}{4a^3}(a^2-x^2)
   +\frac{4a^5 \dot{a}}{9 A^2}  \left[\frac{x}{a} \ln\frac{a+x}{a-x} + \ln\frac{a^2-x^2}{a^2} - \frac{3 x^2}{2a^2} + \frac{1}{2}\ln\frac{e^3}{16} \right],
   \label{eq:houter}
\end{align}
retaining leading- and first-order terms. For matching to the inner region, the key variable is the cube of the slope, and using \eqref{eq:houter} as $x\to a$ (since the ridge is now symmetric, we consider only the right contact line) we find
\begin{align}
   -\left(\frac{\partial h}{\partial x}\right)^3 = \frac{27 A^3}{8 a^6} + 3 \dot{a} \ln\frac{e^3 (a - x)}{2a} + \ldots\,.
   \label{eq:matchout}
\end{align}
We then consider an inner region near the contact line, where $h=O(\lambda)$ and $a-x = O(\lambda)$. We again follow \cite{sibley2015comparison} up until the use of the contact angle boundary condition, that is slightly different here. The slope behaviour for matching from the inner region (i.e.~as $[(a-x)/\lambda] \to \infty$) is thus
\begin{align}
    -\left(\frac{\partial h}{\partial x}\right)^3 
 = C_a^3 + 3 \dot{a}\left[\ln\frac{a - x}{\lambda} + 1 + \ln C_a + C_a^2C_b\right] + \ldots,
   \label{eq:matchin}
\end{align}
where $C_a$ and $C_b$ arise from integration and are fixed by the contact angle condition. To use this, we note that the contact line in the inner region is still located at $x=a$, and we have the behaviour
\begin{align}
    h(x\to a) = C_a ({a - x}) + \dot{a} C_b ({a - x}) + \ldots,
\end{align}
such that
\begin{align}
\label{eq:innerslope}
    \theta = -\left.\frac{\partial h}{\partial x}\right|_{x\to a} =
    C_a + \dot{a} C_b  + \ldots.
\end{align}
The contact angle condition \eqref{basympsqrtLW} is
\begin{align}
\label{eq:CAcondUse}
    \dot{a} &= -\frac{1}{2}M_{\bm{\Gamma}}^0 h_x|_{x=a}[(h_x|_{x=a})^2-\theta_{\mathrm{s}}^2] ,
\end{align}
so using \eqref{eq:innerslope} we have
\begin{align}
    \dot{a} &= \frac{1}{2}M_{\bm{\Gamma}}^0 (C_a + \dot{a}C_b)[(C_a^2 + 2\dot{a}C_aC_b)-\theta_{\mathrm{s}}^2] + O(\dot{a}^2).
\end{align}
To make progress, we assume $\dot{a}\ll M_{\bm{\Gamma}}^0 \ll \dot{a}^{-1}$.
At $O(\dot{a}^0)$, we have $C_a^3 = C_a \theta_{\mathrm{s}}^2$, so that
$C_a = \theta_{\mathrm{s}}$, and $O(\dot{a})$ the expression then simplifies to
\begin{align}
    C_b &= \frac{1}{M_{\bm{\Gamma}}^0 \theta_{\mathrm{s}}^2}.
\end{align}
Using these expressions for $C_a$, $C_b$ and equating the two matching behaviours \eqref{eq:matchout}--\eqref{eq:matchin}, we finally find the equation for the spreading radius $a$ as
\begin{align}
    \dot{a} = \frac{(27 A^3)/(8 a^6) - \theta_{\mathrm{s}}^3}{3\left[
    -2 + \ln(2 a \theta_{\mathrm{s}}/\lambda) + (1/M_{\bm{\Gamma}}^0)
    \right]}.
    \label{eq:asympa}
\end{align}
Equation \eqref{eq:asympa} is a simple first order ODE for $a$ that is easily solved numerically by any ODE solver, with a result for the ridge width $X_+-X_- = 2a$.
It reassuringly has the property that $\dot{a}\to0$ as the slip length $\lambda\to0$.
The numerical solution of \eqref{eq:asympa} gives $a(t)$ and $\dot{a}(t)$, which with a rearranged Eq.~\eqref{eq:CAcondUse}, enables the corresponding time evolution of the contact angle $\theta = -h_x|_{x=a}$ to be found.

Equation~\eqref{eq:asympa} is effectively the well-known Cox-Voinov law \citep{Cox_1986,voinov1976hydrodynamics,snoeijer2013moving}, with $\dot{a}$ taking the place of the Capillary number and the first term in the numerator on the right hand side being the cube of the apparent contact angle---see the leading order of \eqref{eq:matchout}. It is also consistent with Tanner's law \citep{tanner1979spreading} at times away from equilibrium, where $a = (ct)^{1/7}$, albeit with $c$ having a weak logarithmic dependence of
time, consistent with other models \citep{sibley2012pof,sibley2015comparison}.

Whilst Eq.~\eqref{eq:asympa} is simple to solve, the procedure to obtain it is non-trivial. Some earlier authors investigating similar problems with dynamic contact angle conditions instead employed the following na\"ive calculation: Find the static profile of the ridge at a given width (equivalent to our leading order quasi-static outer solution, \eqref{eq:parabola}), and assume it holds at the contact line so that its slope can be used in the expression for $\theta$ in \eqref{basympsqrtLW} to obtain the contact line velocity. In the equivalent notation of \eqref{eq:asympa}, this much simpler procedure gives
\begin{align}
    \dot{a} = \frac{1}{2}M_{\bm{\Gamma}}^0\frac{3A}{2a^2}\left[
    \left(\frac{3A}{2a^2}\right)^2 - \theta_\mathrm{s}^2
    \right].
    \label{eq:naivea}
\end{align}

\begin{figure}
    \centering
    \includegraphics[width=0.99\linewidth]{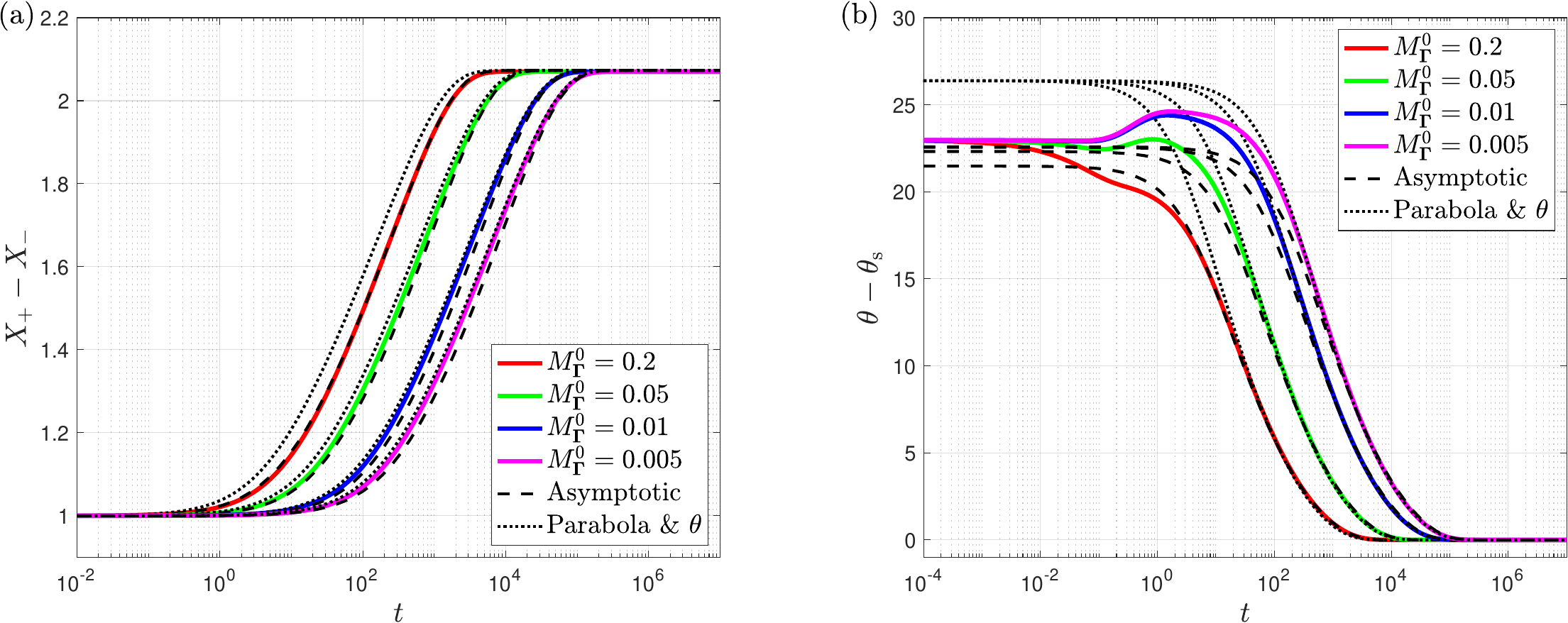}
    \caption{Two-dimensional liquid ridge spreading on a homogeneous surface without gravity. Comparison between (i) asymptotic results exploiting the thin (i.e.~long-wave approximation $|h_x|\ll1$), slow ($|\dot{X}|\ll1$) and small-slip ($\lambda\ll1$) scenario (black dashes), and (ii) a na\"ive evolution formed from using the equivalent static (parabolic) profile in the dynamic contact angle condition (black dots), against (iii) the full numerical solution for equations  \eqref{hasympsqrt}--\eqref{basympsqrt} (solid colours). Panel (a) shows the evolution of the ridge base length $X_+-X_-$ for four values of the surface mobility coefficient, $M_{\bm{\Gamma}}^0 = 0.005$, 0.01, 0.05 and 0.2. Panel
    (b) shows the equivalent results for the contact angle difference $\theta - \theta_\mathrm{s}$. Results shown are for $\theta_\mathrm{s} = 8^\circ$, $\lambda=0.01$, and $A=0.1$. For all lines initial ridge profiles at $t=0$ have half-width $a=0.5$ (equivalently $X_+-X_-=1$), and for the full-numerical solutions the initial profiles are chosen as the parabola in \eqref{eq:parabola}.}
    \label{fig:asymptotic}
\end{figure}

Figure~\ref{fig:asymptotic} shows the comparison between the asymptotic result of \eqref{eq:asympa}, the na\"ive evolution of \eqref{eq:naivea} and a full numerical solution of equations \eqref{hasympsqrt}--\eqref{basympsqrt}, see Appendix~\ref{app:num_ridge} for details of the numerical scheme. Since the full numerical solution does not exploit the long-wave approximation, we choose a relatively small equilibrium contact angle of $\theta_\mathrm{s} = 8^\circ$ to enable a comparison. Similarly, we choose $\lambda=0.01$ which is small enough for the asymptotic procedure, although moderately large compared to physical values.
Four values of the contact angle mobility coefficient $M_{\bm{\Gamma}}^0 = 0.2$, 0.05, 0.01 and 0.005 are shown, with the corresponding asymptotic and na\"ive evolution results given in each case by black dashed and dotted lines, respectively. Panel (a) of Fig.~\ref{fig:asymptotic} shows the evolution of the ridge width, while panel (b) shows the equivalent results for the contact angle difference $\theta - \theta_\mathrm{s}$. The initial condition for the full numerical solution is chosen as the parabola in \eqref{eq:parabola} with $A=0.1$ and $a=0.5$ (equivalently $X_+-X_-=1$).
All three solution methods start from the same ridge width.
The na\"ive solutions have the same initial $\theta$ as each other as they all satisfy $\theta = - h_x|_{x=a} = 3A/(2a^2)$ initially, whereas the asymptotic solution necessarily starts from different values of $\theta$, given it is found using \eqref{eq:asympa}, which depends on $M_{\bm{\Gamma}}^0$.
The full numerical solutions all start at the same $\theta$ but not at the same value as the na\"ive solution (even though both are based on the same initial parabola) because the full numerical solution computes the angle without making the long-wave approximation\footnote{Various alternative ways to compute the angle $\theta$ could straightforwardly be investigated, although outside the scope of the present work.
For instance, whilst not consistent with the asymptotic procedure or the quadratic $h$ profile, the contact angle in all cases could be computed via $\cos\theta = 1/\sqrt{1+(h_x|_{x=a})^2}$. Also, given the full numerical procedure has some flexibility with its initial condition, another option is that a quartic profile could be used that enables symmetry through $h_x|_{x=0}=h_{xxx}|_{x=0}=0$, and has all the desired $\theta_s$, $a$ and
$A$. Finally, an option for the na\"ive approximation is to use the full (not long-wave) static profile, which is the arc of the circle $x^2 + (h + a/\tan\theta)^2 = a^2 \csc^2\theta$ with cross-sectional area $A=a^2(\theta\csc^2\theta - \cot\theta)$.}, i.e.~via $\cos\theta = 1/\sqrt{1+(h_x|_{x=a})^2}$. The over/under-shooting of the contact angle in the full-curvature numerical solutions, when compared to long-wave results, at intermediate times is a result of the profile relaxing from the chosen parabolic initial condition to a profile closer to the arc of a circle. This could be avoided with a different initial condition, but a discrepancy will always occur in one of $A$, $a$, $\theta$ or through relaxation to a slightly different shape. We also note that the contact line velocity $\dot{X_\pm}$ (i.e.~$\dot{a}$, or $\dot{a}/M_{\bm{\Gamma}}^0$ for a fair comparison) could be shown for each line in Fig.~\ref{fig:asymptotic}, but the resulting behaviour is essentially the same as the contact angle evolution due to their close link---c.f.~Eqs.~\eqref{basympsqrt} and \eqref{basympsqrtLW}.

The agreement shown in Fig.~\ref{fig:asymptotic} is sufficient to validate our numerical procedure, but it is noteworthy how well the na\"ive evolution \eqref{eq:naivea} performs. This na\"ive evolution does best for small $M_{\bm{\Gamma}}^0$ where the motion is slowest, which is to be expected, given the result requires the static profile for a given ridge width to give an approximation for the moving contact angle.
However, given the evolution in \eqref{eq:naivea} does not depend on $\lambda$, it is probably more to do with the form of the contact angle condition being similar to the careful asymptotic matching than lending confidence to the na\"ive procedure in general.

\subsection{Liquid ridges on rough surfaces}
\label{subsec:inclined}

Another simple case worth considering is an extension of the previous one, where the substrate is inclined relative to the horizontal and exhibits two-dimensional corrugations.
Again, we assume the liquid thickness is restricted to only vary in the $x$-direction, but now we assume the solid height is also varying in the $x$-direction.
Consequently, $h(\mathbf{r},t)$ and $\phi(\mathbf{r})$ reduce to $h(x,t)$ and $\phi(x)$, respectively. The free energy reads as:
\begin{equation}
\label{Fridge}
\frac{F}{L} = \int_{X_-}^{X^+} \left\{ \sqrt{1+\xi_x^2} - \PP\,\sqrt{1+\phi_x^2} + \frac{B}{2}\left( \xi^2-\phi^2 \right) \right\}\,dx \,.
\end{equation}
 The dynamic contact angle can be related to the slope of $\phi(x)$ and $\xi(x,t)$ at the terminal points as follows. Let $\alpha_{\pm}(t)$ and $\beta_\pm(t)$ be the angles between $z=0$ and $z=\phi(x)$, and $z=0$ and $z=\xi(x,t)$ at $x=X_\pm(t)$, respectively. Accordingly, $\mp\,\tan(\alpha_\pm(t)) = (\phi_x)_{x=X_\pm(t)}$ and $\mp\,\tan(\beta_\pm(t)) = (\xi_x)_{x=X_\pm(t)}$, while the contact angles can be expressed as
\begin{equation}
\label{eq:beta_alpha}
\theta_\pm(t) = \beta_\pm(t) - \alpha_\pm(t) \,.
\end{equation}
Furthermore, equations (\ref{hdyn}) and (\ref{bdyn}) reduce to:
\begin{eqnarray}
\label{hdynridge} \dot{h} &=& \frac{\partial}{\partial x} \left\{ M_v\,\frac{\partial}{\partial x} \left[ B\,\xi - \frac{\partial}{\partial x} \left( \frac{\xi_x}{\sqrt{1+\xi_x^2}} \right)\right] \right\} \\
\label{bdynridge} \dot{X}_{\pm} &=& \mp\,M_{\bm{\Gamma}}\,\cos(\alpha_\pm)\,(\cos(\theta_\pm)-\PP) \,,
\end{eqnarray}
where (as in the previous case) the change of sign on the right-hand side in Eq.~(\ref{bdynridge}) is due to the fact that $\mathbf{u}=(1,0)$ on $x=X_+(t)$ and $\mathbf{u}=(-1,0)$ on $x=X_-(t)$. Finally, the dynamic boundary condition can be obtained in this geometry using Eq.~\eqref{Fridge} in Eq.~(\ref{DBCh}), thus yielding
\begin{equation}
\label{DBCridge}
    \frac{\dot{X}_{\pm}}{\lambda^2} = \cos^3(\beta_\pm)\, \left[ \xi_{xxx} \pm\,\frac{3}{2}\,\sin(2\,\beta_\pm)\,\xi_{xx}^2 \right] - B\,\xi_x \,,
\end{equation}
where the derivatives $\xi_x$, $\xi_{xx}$, and $\xi_{xxx}$ are evaluated at $x=X_\pm(t)$.

\begin{figure}
    \centering
        \includegraphics[width=0.49\linewidth]{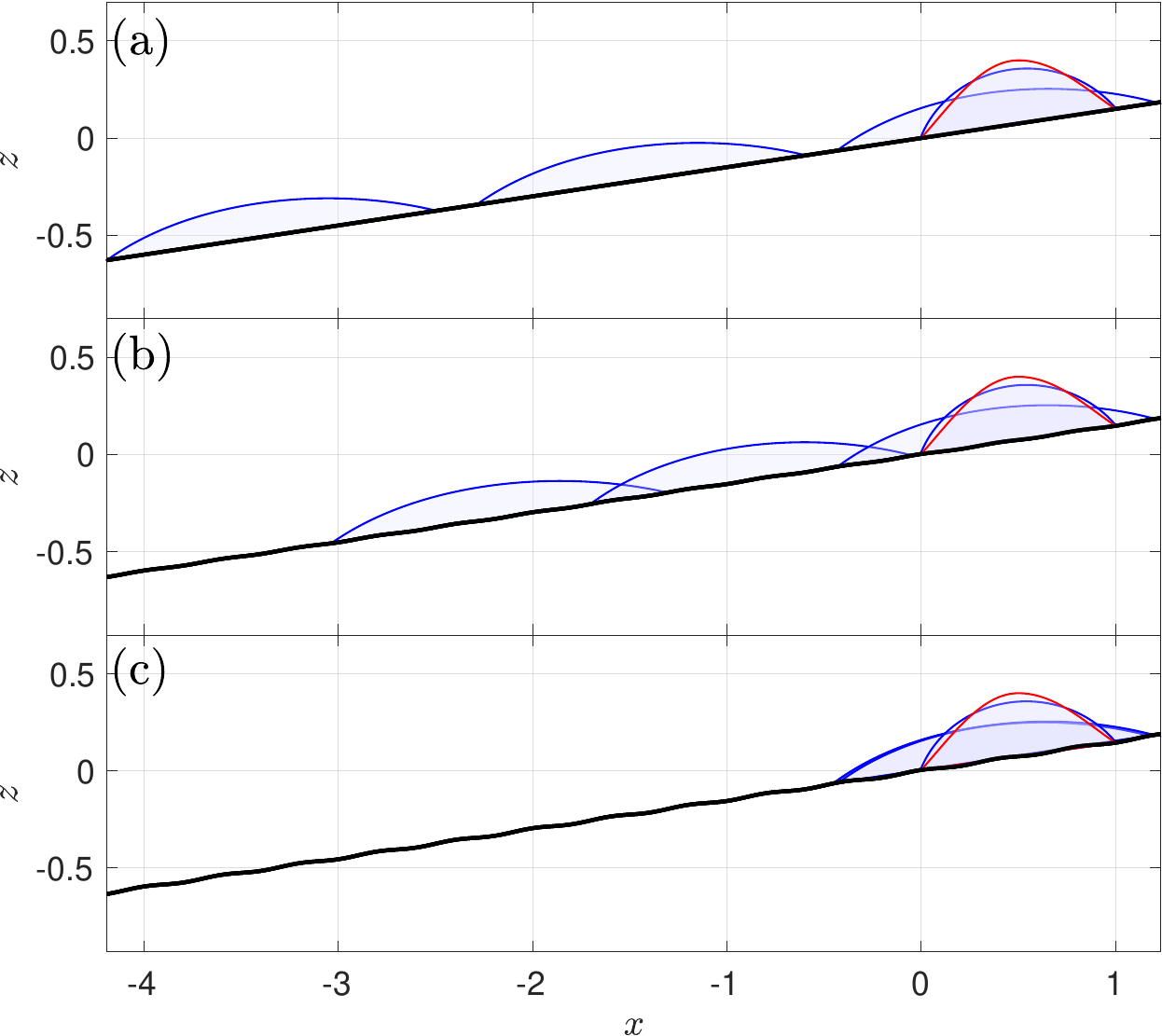}\;
        \includegraphics[width=0.49\linewidth]{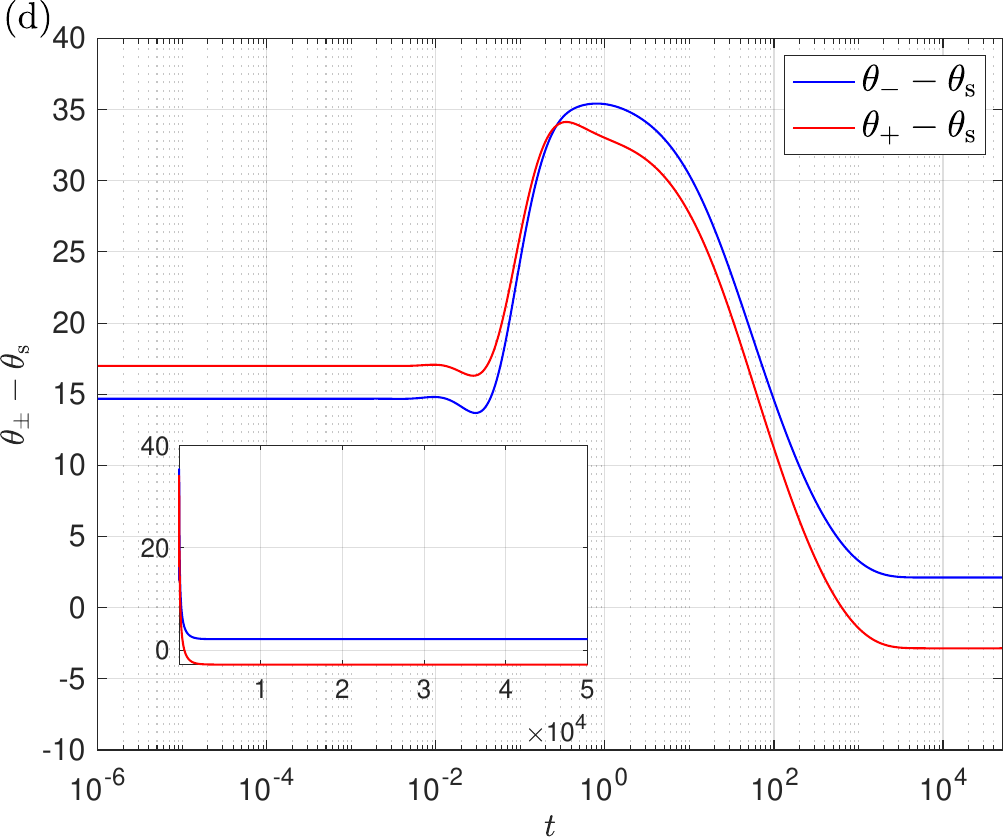}
        \\ \vspace{0.1cm}
        \includegraphics[width=0.49\linewidth]{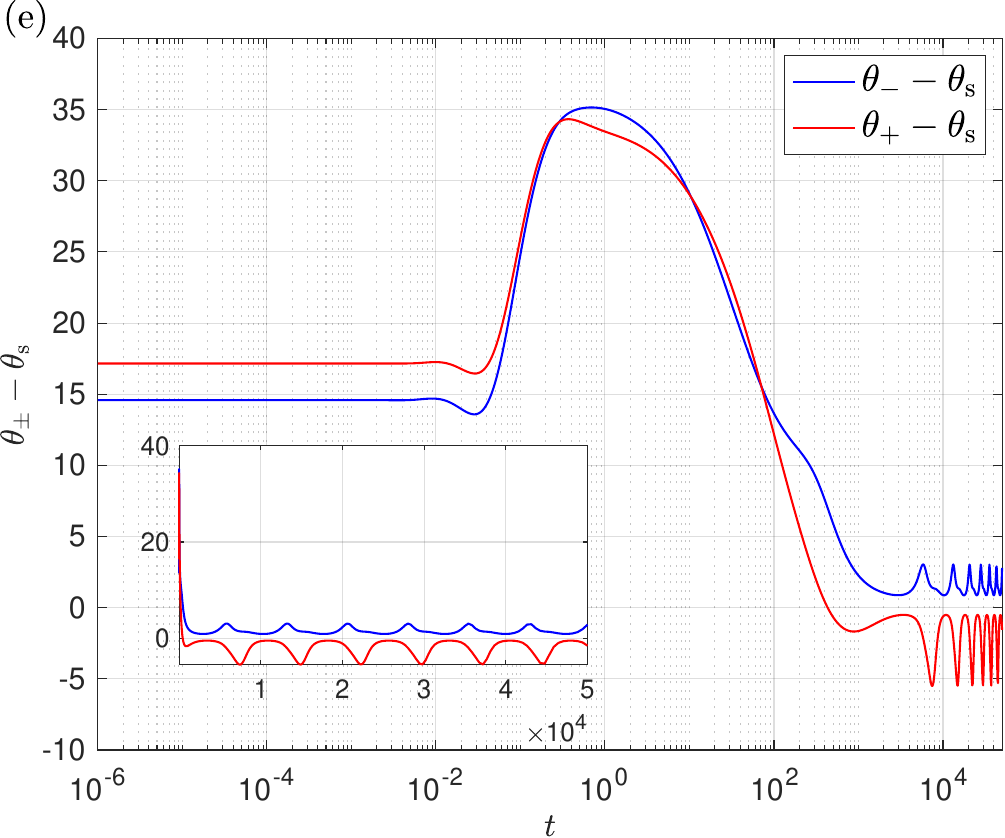}\;
        \includegraphics[width=0.49\linewidth]{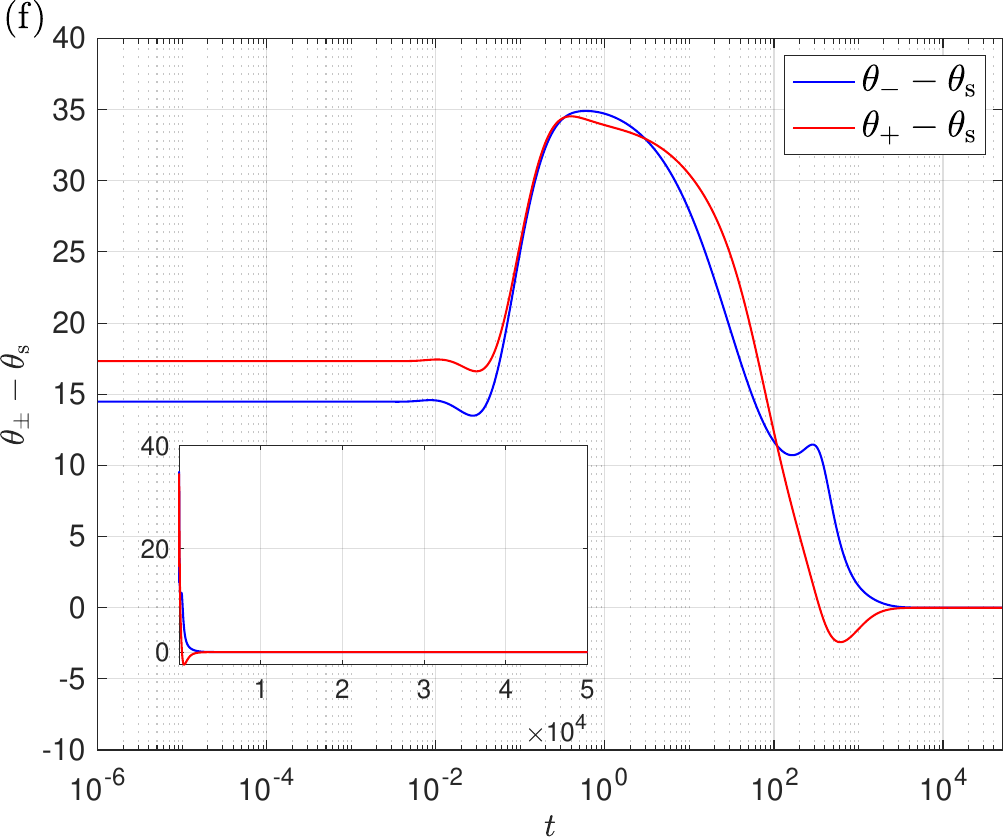}
    \caption{Ridge motion down inclined corrugated surfaces. The equilibrium contact angle $\theta_s = 25^\circ$, the slip length $\lambda=0.01$, the surface mobility coefficient $M_{\bm{\Gamma}}^0=0.01$ [see Eq.~\eqref{eq:contact_line_mob}] and the Bond number $B=1.25$. The substrate surface profiles are given in Eq.~\eqref{eq:surface}, with inclination slope of 0.15 and oscillations with wavelength 2/5 and amplitudes of (a,d) $A=0$; (b,e) $A=0.0025$; (c,f) $A=0.005$. Initial conditions ($t=0$) are shown as the red profiles in (a--c) with ridge cross-sectional area of 0.203 and subsequent (blue) profiles are at the times $t=1$, 1000, 25000 and 50000.}
    \label{fig:3fig}
\end{figure}

An illustrative situation that can be described by Eqs.~\eqref{hdynridge}--\eqref{DBCridge} is that of a droplet (ridge) sliding down an inclined corrugated surface.
The overall phase behaviour of liquids on corrugated surfaces is very rich \citep{malijevsky2025complete}.
Thin-film models for liquids flowing over corrugated substrates (including those with various external effects, such as an applied electric field) have been previously derived from the full governing equations using systematic asymptotics \citep{stillwagon1990leveling, tseluiko2008electrified, tseluiko2011electrified, tseluiko2013stability} and the dynamics of droplets/ridges on such surfaces has previously been considered \citep{savva2009two, savva2011contact_pt1, savva2011contact_pt2, savva2011dynamics, savva2012influence, savva2013droplet, vrionis2025efficient}.
So we restrict ourselves to one illustrative case, that of a ridge of liquid (assumed to be varying only in the $x$-direction) on an inclined corrugated surface with height profile
\begin{equation}\label{eq:surface}
    \phi(x) = A\cos(5\pi x) + 0.15x,
\end{equation}
corresponding to a slope with inclination gradient 0.15 and with the amplitude of the corrugations $A$.
Such corrugations can be manufactured, for example, via the repeated deposition of coffee rings, as described e.g.\ in \cite{thiele2014patterned, shawail2025tunable}.
Note that although Eq.~\eqref{hdynridge} has a form similar to previously derived thin-film models for liquids flowing over corrugated surfaces (see \cite{stillwagon1990leveling, tseluiko2008electrified, tseluiko2011electrified, tseluiko2013stability}), it is not fully equivalent to those models.
In contrast to earlier studies, where reduced-curvature approximations were typically employed, our formulation incorporates the full curvature.
As an aside, note also previous works usually choose the coordinate system such that the $x$-axis aligns with the mean flow direction, whereas this is not the case here.

In Fig.~\ref{fig:3fig} we display results for three different values of the corrugation amplitude $A$.
Panel (a) presents results for $A=0$ (smooth surface), showing the ridge spreading from the initial $t=0$ condition (red line) and then sliding to the left, down the slope.
The blue curves correspond to the ridge profile at the times $t=1$, 1000, 25000 and 50000.
Panel (b) presents the corresponding results for corrugation amplitude $A=0.0025$, showing the ridge still sliding down, but at a slower velocity, so it does not travel as far to the left at each of the times.
Panel (c) shows the case when $A=0.005$.
This corrugation amplitude is large enough that the liquid ridge becomes pinned on the rough surface.
The de-pinning transition that occurs as $A$ is decreased can also be observed for fixed $A$ by increasing the droplet volume.
This has been studied previously using precursor-film models, with the surface heterogeneity treated as being chemical in origin \citep{beltrame2011rayleigh, engelnkemper2019collective}, leading to results that are qualitatively inline with the results presented here.

In Fig.~\ref{fig:3fig}(d)-(e) we display plots of the contact angle difference $\theta_\pm-\theta_\mathrm{s}$ of the ridge over (log) time.
The inset in each case shows the same data plotted over time (no logarithmic axes).
$\theta_-$ is the down-hill dynamic contact angle, $\theta_+$ is the up-hill one and $\theta_\mathrm{s}=25^\circ$ is the equilibrium contact angle.
Figure~\ref{fig:3fig}(d) shows the case when $A=0$, where after a time $t\approx3000$ the ridge settles into a shape that is stationary in the centre-of-mass reference frame, with the two dynamic contact angles $\theta_\pm\neq\theta_\mathrm{s}$.
Figure~\ref{fig:3fig}(e) shows the case when $A=0.0025$, where after a time $t\approx10000$ the ridge settles into a state that in the centre-of-mass reference frame corresponds to a periodic orbit, as the ridge slides past each subsequent maximum on the surface.
This can best be seen in the inset, which highlights better the long-time behaviour, showing the periodic behaviour of the dynamic contact angles $\theta_\pm$.
Finally, Fig.~\ref{fig:3fig}(e) shows the case when $A=0.005$, where after a time $t\approx3000$ the ridge settles into a stationary state with the contact angles equal to the equilibrium contact angle, $\theta_\pm=\theta_\mathrm{s}$, as it becomes pinned on the surface.
When comparing Figs.~\ref{fig:3fig}(d--f) it is important to note that it is necessary in our model (in particular in the dynamic contact angle condition) that $\theta_\pm=\theta_\mathrm{s}$ for a droplet to be static. Throughout this work when referring to the contact angle, we describe the \textit{microscopic} contact angle at the surface, whereas an \textit{apparent} angle (an extrapolated or measured angle on the macroscale of the overall droplet profile) could exhibit asymmetric bending (and thus $\theta_+\neq\theta_-$) due to gravity \citep{bonn2009wetting}.

\subsection{Axisymmetric droplet spreading}
\label{subsec:axisymmetric}

\begin{figure}
    \centering
        \includegraphics[width=0.8\linewidth]{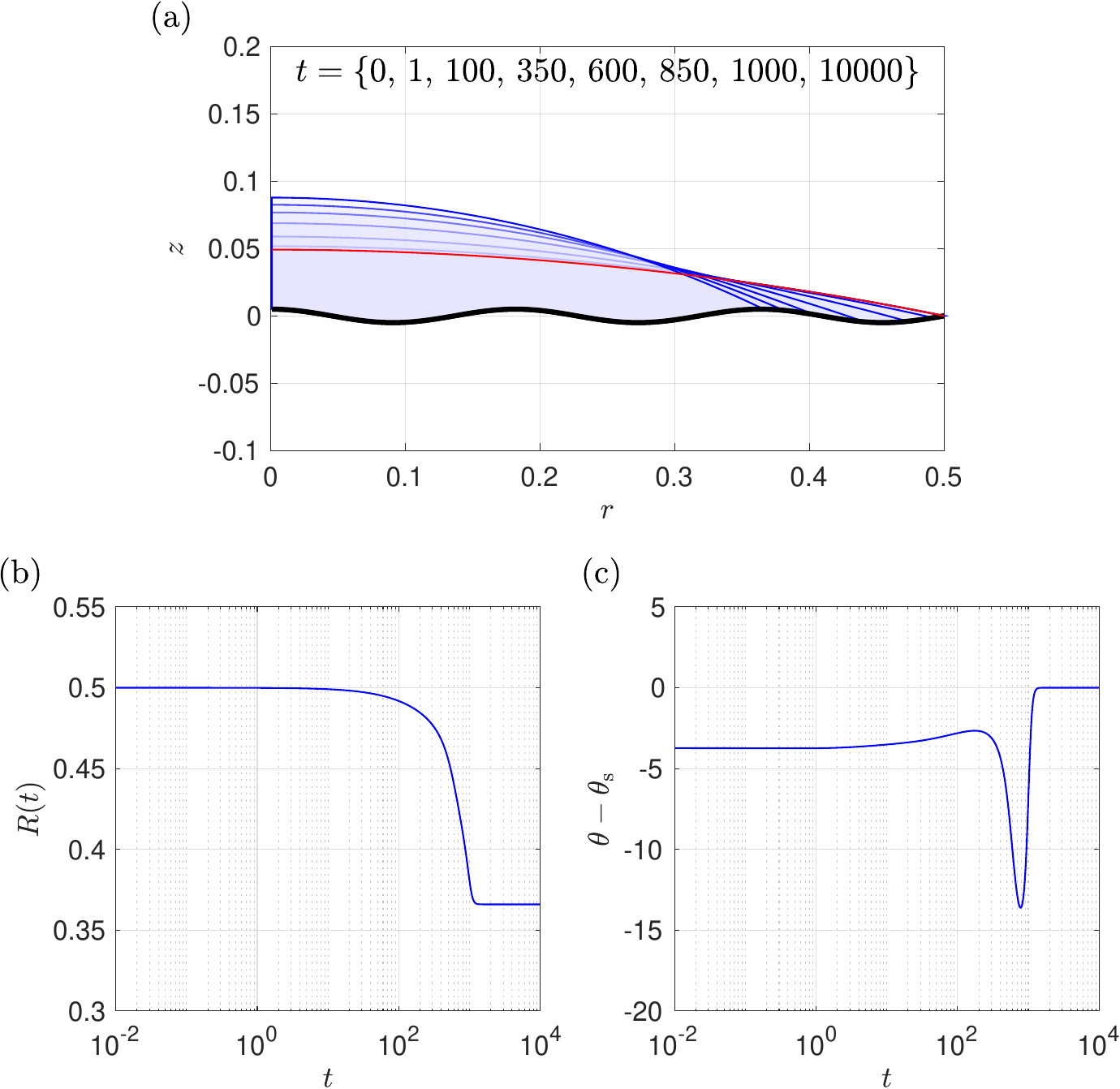}
    \caption{Axisymmetric droplet motion on surface with sinusoidal height profile (also axisymmetric). Panel (a) shows the $t=0$ initial condition (red profile) together with the subsequent drop profile over time (blue profiles), for the times $t=1$, 100, 350, 600, 850, 1000 and 10000. Panel (b) shows the drop radius over time, while (c) shows the contact angle difference $\theta-\theta_\mathrm{s}$ over time. The equilibrium contact angle $\theta_s = 25^\circ$, the slip length $\lambda=0.01$, the contact line mobility $M_{\bm{\Gamma}}^0=0.01$ and the Bond number $B=1.25$. The surface height profile is $\phi(r) = 0.005\cos(11\pi r)$. The drop volume is 0.02.}
    \label{fig:figaxi}
\end{figure}

Besides the liquid ridge, another physically relevant scenario is when the solid and the liquid are assumed to be axisymmeric \citep{Hocking83}.
In this case, the liquid thickness and the solid surface are described by $h(r,t)$ and $\phi(r)$, respectively, where $r=\sqrt{x^2+y^2}$ is the distance from the centre of the drop.
The boundary of the domain is a circle with time-dependent radius $R(t)$.
The free energy is then written as
\begin{equation}
\label{Fcircle}
F = \int_{0}^{R} \left\{ \sqrt{1+\xi_r^2} - \PP\,\sqrt{1+\phi_r^2} + \frac{B}{2}\left( \xi^2-\phi^2 \right) \right\}\,(2\pi\,r)\,dr \,.
\end{equation}
By expressing the functional derivative in polar coordinates, the evolution equation for the droplet height becomes
\begin{equation}
\label{axialht} \dot{h} = \frac{1}{r}\,\frac{\partial}{\partial r}\,\left\{ r\,M_v\,\frac{\partial}{\partial r} \left[ B\,\xi - \frac{1}{r}\,\frac{\partial}{\partial r} \left( \frac{r\,\xi_r}{\sqrt{1+\xi_r^2}} \right) \right] \right\} \,.
\end{equation}
The domain boundary can be written in the form of $\mathbf{R}(\chi,t)=R(t)\,\mathbf{n}(\chi)$, where $\mathbf{n}(\chi) = (\cos(2\pi\,\chi),\sin(2\pi\,\chi))$ is the radial unit vector. Consequently, $\mathbf{u}=\mathbf{n}_\perp/s=\mathbf{n}(\chi)$.
Expressing the matrix elements in terms of polar coordinates in Eq.~\eqref{eq:Tdotu} then yields $\mathbb{T}\cdot\mathbf{u} = |\mathbf{n}_\phi|\,[\cos\theta-\PP]\,\mathbf{n}$, where $\theta(t)$ is the dynamic contact angle satisfying $\tan(\alpha(t)+\theta(t))=-(\xi_r)|_{\partial\Gamma}$, where $\tan(\alpha(t))=-(\phi_r)|_{\partial\Gamma}$.
See Eq.~\eqref{eq:beta_alpha} and the corresponding discussion for the definitions of these angles.
Furthermore, using $\mathbb{S} = \left(\mathbb{I} + \phi_r^2\,(\mathbf{n} \otimes \mathbf{n})\right)|_{\partial\Gamma}$ results in $\mathbb{S}^{-1}\cdot\mathbb{T}\cdot\mathbf{u} = \cos(\alpha)\,[\cos\theta-\PP]\,\mathbf{n}$. Using this result together with $\dot{\mathbf{R}}=\dot{R}\,\mathbf{n}$ in Eq.~(\ref{bdyn}) yields:
\begin{equation}
\label{axialRt} \dot{R} = - M_{\bm{\Gamma}}\,\cos(\alpha)\,\left[ \cos\theta-\PP \right] \,.
\end{equation}
The dynamic boundary condition reads as:
\begin{equation}
\label{DBCax}
    \frac{\dot{R}}{\lambda^2} = \cos^3(\beta)\,\Gamma(\xi_r,\xi_{rr},\xi_{rrr}) -B\,\xi_r + \frac{\sin\beta}{R^2} \,,
\end{equation}
where $\beta(t)=\alpha(t)+\theta(t)$, and 
\begin{equation}
    \Gamma(\xi_r,\xi_{rr},\xi_{rrr}) = \xi_{rrr}+\frac{3}{2}\,\sin(2\,\beta)\,\xi^2_{rr}+\frac{\xi_{rr}}{R}\,,
\end{equation}
where the derivatives $\xi_r$, $\xi_{rr}$, and $\xi_{rrr}$ are evaluated at $r=R(t)$. Finally, the boundary conditions at $r=0$ are 
\begin{equation}
\label{BCax} h_r|_{r=0}=h_{rrr}|_{r=0}=0 \, ,
\end{equation}
which enforce the profile symmetry. A numerical scheme for solving these equations is described in Appendix~\ref{app:num_axi}.

In Fig.~\ref{fig:figaxi} we present a prototypical example of droplet motion on an axisymmetric substrate with sinusoidal height profile, for the case when $B=1.25$ and the droplet volume is 0.02.
For the droplet evolution to remain axisymmetric throughout the evolution, the substrate and the initial condition both are also imposed with axisymmetry.
This is a case where the droplet retracts due to its contact angle initially being smaller than the imposed static angle.
Panel (a) shows snapshots of the droplet profile over time, with (b) giving the corresponding evolution of the droplet radius $R(t)$.
Fig.~\ref{fig:figaxi}(c) shows the contact angle evolution, demonstrating the effect of the substrate undulations that cause a sudden reduction in the dynamic contact angle when the contact line moves from the downslope to the upslope.
Eventually the droplet attains its equilibrium shape once the dynamic contact angle reaches the imposed static value, as expected.

\subsection{Evaporation of a symmetric ridge on a flat surface without gravity}
\label{subsec:evap_example}

As a final example, we present results for a liquid ridge (2D droplet) on a flat surface that either shrinks due to evaporation or grows as vapour condenses onto it, depending on the value of $p_G$ in Eq.~\eqref{eq:Omega}, i.e.\ on the value the chemical potential in the surrounding vapour phase.
For simplicity, we assume the surface is flat with $\phi=0$ and that the liquid ridge is small enough that we may neglect gravity, so $B=0$.

From Eq.~\eqref{eq:evap_dyn}, we have
\begin{align}
\dot{h} = \frac{\partial}{\partial x} \left[ -h\,(h^2+\lambda^2) \frac{\partial^2}{\partial x^2} \left( \frac{h_x}{\sqrt{1+h_x^2}} \right) \right] + \sigma,\label{eq:evap1ge}
\end{align}
where the evaporation/condensation flux $\sigma(x,t)$ is
\begin{align}
\sigma = - M_{nc}\left[
-\frac{\partial}{\partial x} \left( \frac{h_x}{\sqrt{1+h_x^2}} \right)-p_G
\right], \label{eq:evap2ge}
\end{align}
and from Eqs.~\eqref{bdyn} and \eqref{BMA_grand} the contact line evolution is given by
\begin{align}
\label{evap3ge} \pm\dot{a} = \mp\,M_{\bm{\Gamma}}^0 \,\theta[\cos\theta-\cos\theta_{\mathrm{s}}] \,,
\end{align}
where $-a\leq x\leq a$ such that $x=\pm a(t)$ are the contact line locations. We have the $\pm$ in front of the $\dot{a}$ in Eq.~\eqref{evap3ge}, to highlight the equal and opposite dynamics of the two contact lines, due to symmetry.

Equation~\eqref{evap3ge} comes from the transversality condition (see Eqs.~\eqref{bdyn} and \eqref{BMA_grand}), so at face value, one might assume it depends on $\omega = f-p_G h$.
However, because it applies at the contact line where $h=0$, the resulting expression for $\dot{a}$ ends up being identical to the result from Eq.~\eqref{bdynridge}, i.e.\ the same as when there is no evaporation.
In contrast, evaporation does enter the dynamic boundary condition \eqref{evapDBCh}, which at $x=a$ gives
\begin{align}
    h_x|_{x=a}\left\{ -\lambda^2\,\frac{\partial^2}{\partial x^2} \left( \frac{h_x}{\sqrt{1+h_x^2}} \right)\bigg|_{x=a} + \dot{a} \right\}
    =  M_{nc}\left[
-\frac{\partial}{\partial x} \left( \frac{h_x}{\sqrt{1+h_x^2}} \right)-p_G
\right]\bigg|_{x=a}.
\label{eq:dynBCe}
\end{align}
Somewhat motivated by the na\"ive approximation of \S~\ref{sec:liq_ridge}, we compare the full numerical solution of Eqs.~\eqref{eq:evap1ge}--\eqref{eq:dynBCe} 
(using a slightly adapted numerical scheme from Appendix~\ref{app:num_ridge}) against an approximation using the arc of a circle for the height profile $h(x,t)$. This is done partly to validate the numerical scheme.
However, the na\"ive approximation is also of intrinsic interest and value, due to its simplicity.

For zero evaporation/condensation flux $\sigma$, we have at equilibrium that
\begin{equation}
    \frac{\delta \Omega}{\delta h} =  -\frac{\partial}{\partial x} \left( \frac{h_x}{\sqrt{1+h_x^2}} \right) - p_G
\end{equation}
is equal to a constant (that effectively fixes either the volume or the contact angle).
From this, a few steps of algebra shows that the free surface is the arc of a circle satisfying
\begin{equation}
\label{eq:circ_evap}
    x^2 + (h + a\cot\theta)^2 = a^2\csc^2\theta,
\end{equation}
which is (chosen to be) symmetric at $x=0$ with contact lines at $x=\pm a$, i.e.~$h|_{x=\pm a} = 0$ and at contact angle $\theta$. Using this profile, the cross-sectional area is
\begin{align}
    A = a^2(\theta\csc^2\theta - \cot\theta).
    \label{thandA2}
\end{align}
Simple geometry can relate these quantities to one another in various ways. For instance, along with $\tan\theta = \mp h_x|_{x=\pm a}$, it is useful to note that
\begin{align}
    \sin\theta = \mp\left.\frac{h_x}{\sqrt{1+h_x^2}}\right|_{x=\pm a}, \quad
    \cos\theta = \left.\frac{1}{\sqrt{1+h_x^2}}\right|_{x=\pm a}.
\end{align}
Using Eq.~\eqref{eq:circ_evap} in \eqref{eq:evap1ge}--\eqref{evap3ge} we have (by construction) that $\dot{h}=\sigma$, so
\begin{align}
    \dot{h} = - M_{nc}\left[
\frac{\sin\theta}{a}-p_G
\right].
\end{align}
We can also find the evolution of the cross-sectional area,
\begin{align}
    \dot{A} &= \frac{d}{d t}\left(
    \int_{-a}^{a} h \, dx 
    \right)
\nonumber\\
&=  \int_{-a}^{a} \dot{h} \,dx + \dot{a} h|_{x=a} + \dot{a} h|_{x=-a}\nonumber\\
&=\int_{-a}^{a} - M_{nc}\left[
\frac{\sin\theta}{a}-p_G
\right] dx \nonumber\\ &= 2M_{nc}\left(a \, p_G-\sin\theta
\right).\label{eq:dA_dt}
\end{align}
Therefore, Eqs.~\eqref{evap3ge} and \eqref{eq:dA_dt} together form a pair of first order ODEs for $A(t)$ and $a(t)$ in terms of $\theta$ and $a$. Together with Eq.~\eqref{thandA2}, which expresses $A$ in terms of $a$ and $\theta$, these three equations provide a simple approximation for computing the evolution of $A$, $a$ and $\theta$ and any other quantity derived from the $h$ profile.

\begin{figure}
    \centering
    \includegraphics[width=0.99\linewidth]{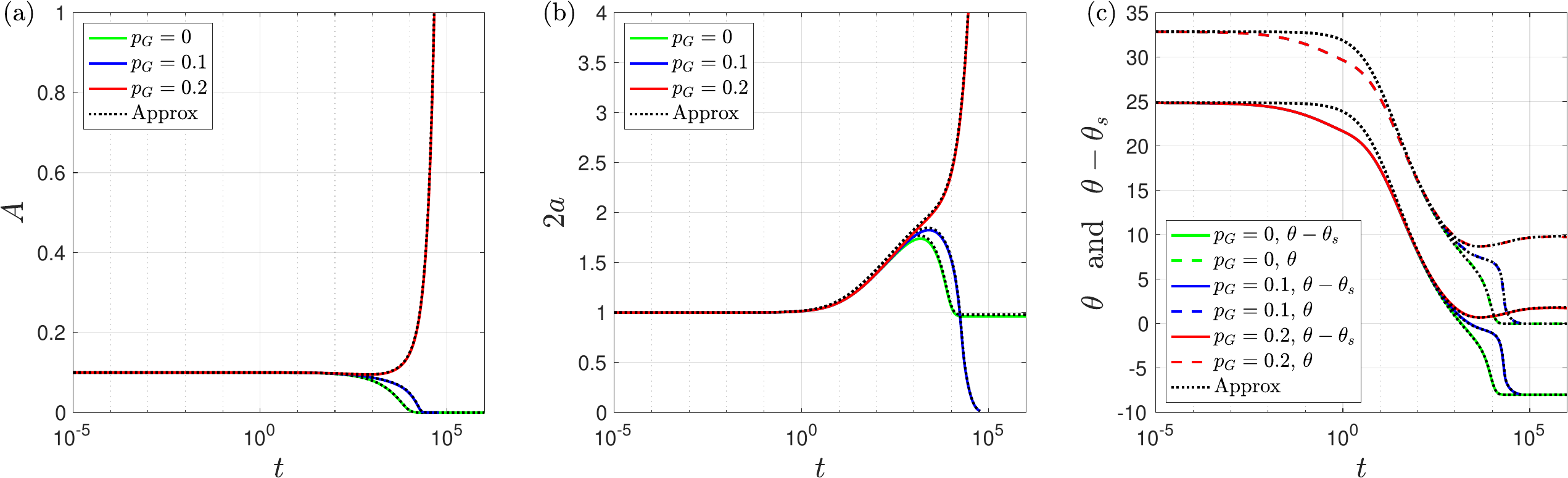}\\\vspace{0.1cm}
    \includegraphics[width=0.99\linewidth]{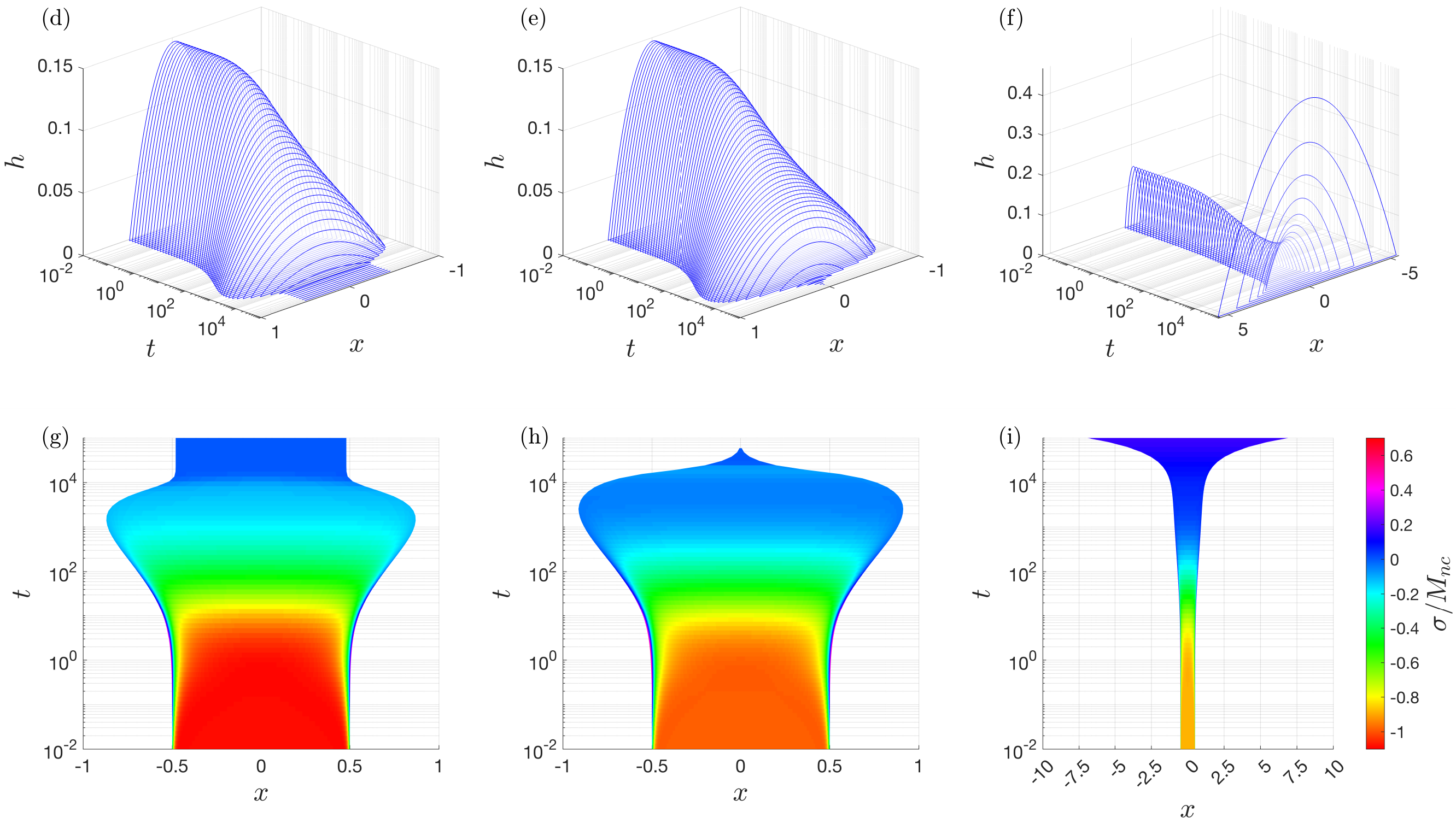}
    \caption{Two-dimensional liquid drop (liquid ridge) experiencing evaporation or condensation on a flat surface, with Bond number $B=0$. Panels (a)--(c) show the variation over time of the cross-sectional area $A$, ridge width $2a$ and the contact angle $\theta$, for three different values of $p_G$. In each case, the black dotted line shows the approximate evolution given by Eqs.~\eqref{evap3ge} and \eqref{eq:dA_dt} with \eqref{thandA2}, which assumes the drop shape is the arc of a circle, and agree well with the full numerical solutions of Eqs.~\eqref{eq:evap1ge}--\eqref{eq:dynBCe}. Panels (d)--(f) show film height $h(x,t)$ profiles from the full numerical simulations corresponding to (a)--(c), with (d) being for $p_G=0$, (e) for $p_G=0.1$ and (f) for $p_G=0.2$. Panels (g)--(i) show heatmap plots of the evaporation/condensation flux $\sigma(x,t)$ for the same cases as (d)--(f).
    In all cases, $\lambda=0.01$, $M_{\bm{\Gamma}}^0=0.1$, $M_{nc}=5\times10^{-5}$ and $\theta_s = 8^\circ$. The initial conditions for the full numerical solutions are chosen as the arc of a circle with $A=0.1$ and $a=0.5$.}
    \label{fig:evap}
\end{figure}

In Fig.~\ref{fig:evap} we present results corresponding to the case where the initial profile is a narrow, symmetric, arc of a circle centred at $x=0$ with cross-sectional area $A=0.1$ and half-width $a=0.5$. Three values of $p_G$ are chosen, covering both evaporating and condensing scenarios, noting that $p_G$ need not be negative for evaporation here, due to Laplace pressure from the curved droplet surface.
All other parameters are fixed to values similar to previous cases in this section (see the figure caption for details).
We set $M_{nc}=5\times10^{-5}$, which is small enough that in all cases the ridge has enough time to initially spread to obtain a shape close to that dictated by the equilibrium contact angle $\theta_\mathrm{s}=8^\circ$, before the evaporation/condensation effects become dominant.
Such an interplay between spreading and evaporation resulting in a maximal drop radius at an intermediate time is consistent with experimental observations (see e.g.~discussion of Fig.~8 of \cite{cazabat2010evaporation}).
For $p_G=0$ (corresponding to the green lines in Fig.~\ref{fig:evap}(a)--(c) and the $h(x,t)$ and $\sigma(x,t)$ profiles displayed in (d), (g), respectively), we observe that evaporation is fast enough that the ridge becomes flattened, with zero volume, i.e.\ $A\to0$, for $a=O(1)$.
In other words, the liquid evaporates before the two contact lines are able to meet at $x=0$.
This is possible because $\theta\to0$ sufficiently fast, and hence also $\dot{a}$, due to the prefactor $\theta$ in Eq.~\eqref{evap3ge}.
By contrast, in the case where $p_G=0.1$ (the blue lines in Fig.~\ref{fig:evap}(a)--(c) and profiles in (e) and (h)), the liquid still evaporates, but it does so more slowly, with $A$ and $a$ both approaching zero together, at the end of the evaporation process.
Finally, in the case where $p_G=0.2$ (corresponding to the red lines in Fig.~\ref{fig:evap}(a)--(c) and profiles in (f) and (i)), the vapour chemical potential is sufficiently high that we observe both $A$ and $a$ increasing at later times, through condensation.
The approximate evolution from Eqs.~\eqref{evap3ge}, \eqref{eq:dA_dt} and \eqref{thandA2}, which assumes the $h$ profile stays as the arc of a circle, shows good agreement for the cross-sectional area (a), the ridge width (b), and the contact angle (c) in all cases, and is shown as black dotted lines in these panels.

\section{Concluding remarks}
\label{sec:conc}

In this paper we have shown how OVP \citep{onsager1931reciprocal_1, groot1964non, doi2011onsager, doi2013soft, doi2019application} may be used as the basis for deriving equations of motion for liquid films and drops over uneven surfaces, including obtaining equations of motion for contact lines---see Eqs.~\eqref{hdyn} and \eqref{bdyn}.
We have also shown how the approach can be extended to incorporate evaporation or condensation, resulting in the non-conserved dynamics in Eq.~\eqref{eq:evap_dyn}.
This equation has previously been extensively used and extended to analyse various evaporative dynamics problems, including deposition of solute on surfaces \citep{samid1998pattern, padmakar1999instability, lyushnin2002fingering, thiele2009modelling, thiele2010thin, frastia2011dynamical, frastia2012modelling}.
The advantage of generating the dynamical equations via OVP is that it builds from the free energy $F$ of the system.
Thus, the correct equilibrium properties of the system are naturally incorporated into the non-equilibrium description.
We therefore expect our work to be a valuable starting point for anyone in the future wanting to formulate equations for (slow) low-Reynolds-number flows on surfaces, needing to incorporate additional physical phenomena.
As long as the additional physics can be described by relevant extra terms in the free energy functional, then OVP provides a straightforward way to incorporating the influence of these on the dynamics also.

Another result that naturally follows from the approach taken here is that the equilibrium condition for drops on surfaces may be written as $\det(\mathbb{T})=0$, i.e.\ the transversality condition \eqref{TRC}.
This, together with the identification that this is an alternate way of stating Young's equation \eqref{Young}, makes \eqref{TRC} a useful result.
Moreover, following the arguments of \cite{bormashenko2009young}, in addition to being able to formulate Young's equation this way, so also the Neumann, Wenzel and Cassie-Baxter equations will all result as such conditions.
Additionally, the analysis in \S~\ref{subsec:contact_angle} showing that contact line speed is given by the very simple result in \eqref{eq:V_magnitude_2}, is also very powerful.
This shows that the contact line speed depends on the contact angle $\theta$ and the constant $M_{\bm{\Gamma}}^0\propto\gamma_{LG}/\eta$.
This result facilitates making easy estimates for contact line speeds and droplet/film flow equilibration times.

The results presented in \S 4 show application of this general approach to various limiting cases. Consistency with known results such as the Cox-Voinov law for non-volatile droplet spreading, and qualitative comparison with experimental observations for the interplay between spreading and evaporation were described. The retention of full-curvature in surface-tension driven flows, and the effect of the contact line speed/contact angle relation have also been investigated.

The present work also highlights the effectiveness of OVP \citep{onsager1931reciprocal_1, groot1964non, doi2011onsager, doi2013soft, doi2019application} as a general framework for generating the dynamical equations for overdamped slow (fluid-dynamical) systems.
For systems of interacting Brownian particles, power functional theory \citep{schmidt2013power, schmidt2022power} shows that the Onsager minimisation principle is a formally exact way of generating the dynamical equations, albeit, in general, neither the free energy functional $F$ nor the dissipation functional $\Phi$ are known exactly.
This leads us to take the view that one should take a similar attitude to the theory more generally, i.e.\ that for overdamped dissipative systems Eq.~\eqref{Rayleigh} is good place to start and the goal should be to find good approximations for $F$ and $\Phi$.
In this endeavour it is worth noting some of the broader applications of OVP, such as to active matter \citep{krinninger2019power, wang2021onsager} and to modelling bendotaxis in a thermodynamically consistent manner \citep{ZhQi2022prf}.
Various other broader connections are made in \cite{peng2019conservation, mielke2014relation, MielkeRengerPeletier2016}.
As \cite{te2024microscopic} shows, memory effects may be relevant in certain situations.
To go beyond the quadratic dissipation functionals that we assume here in Eqs.~\eqref{eq:Phi} or \eqref{eq:Phi_grand}, on the basis of \cite{te2024microscopic}, one may expect the corresponding $\Phi$ to in some cases involve a memory function and a corresponding time integral over the history of the system dynamics (c.f.\ also \cite{schmidt2013power}).
We believe that future investigations along these lines will be fruitful.

\section*{Acknowledgements}

We are grateful to Ricardo Barros, Marco Mazza, Harry Scott and Uwe Thiele for valuable input and discussions on many of the ideas here. We also thank the anonymous referees for their helpful comments.

\noindent {\bf Funding.} A.J.B.-T.\ was supported by an EPSRC studentship.

\noindent {\bf Declaration of Interests.}
The authors report no conflict of interest.

\appendix

\section{Dissipation functional}
\label{app:df}

For readers who are unfamiliar with the dissipation functional, we recommend the broad general overviews given by \cite{doi2011onsager, doi2013soft}.
In the context of thin liquid films on surfaces, the following arguments which build on \citep{qian2006variational, lopes2018multiple, peschka2018variational, doi2021onsager}, provide useful context on the origin of the first term in \eqref{eq:Phi}.
Consider a thin film of liquid on a horizontal flat surface; to simplify the argument, we assume the height $h$ is varying only in the direction of the $x$-axis (see Figure~\ref{fig:drop_sketch}).
We first discuss the dissipation in the body of the droplet/film and discuss the contact line dissipation separately, below.
The viscous dissipation within the liquid film due to gradients in the velocity is
\begin{equation}\label{eq:Phi_appendixB}
    \Phi=\frac{\eta}{2}\int \left[\frac{h(u|_{z=0})^2}{\lambda^2} + \int_0^h \left(\frac{\partial u}{\partial z}\right)^2 dz\right]  dx,
\end{equation}
where $u(z,t)$ is the velocity of the fluid within the film, parallel to the surface\footnote{
To compare more directly to \cite{peschka2018variational}, we note that the first term in the integral of Eq.~\eqref{eq:Phi_appendixB} could instead be written more generally as
 $ (1/\Lambda) (u|_{z=0})^2$,
where $\Lambda$ depends on the slip law via $u|_{z=0} =\Lambda u_z|_{z=0}$, i.e.~for Navier-slip it would be $\Lambda = \lambda$ and for our chosen slip condition it is $\Lambda = (\lambda^{2}/h)$. The second term comes from the incompressible Newtonian fluid / Stokes flow dissipation $\int\frac{1}{4\eta} \bm{\tau}:\bm{\tau} \, dV = \int \frac12 \bm{\tau}:\bm{\mathbb{D}} \, dV = \int \frac12 \bm{\tau}:(\nabla \mathbf{v}) \, dV$, where $\bm\tau = 2\eta\bm{\mathbb{D}}$ is the Newtonian viscous stress tensor and $\bm{\mathbb{D}} = \frac12(\nabla \mathbf{v} + (\nabla \mathbf{v})^T)$ is the rate of strain tensor. In two-dimensional flow (with spatial coordinates $(x,z)$ and $\mathbf{v}=(u,v)$), this becomes $\int \frac{\eta}{2} (2u_x^2 + (u_z+v_x)^2 + 2v_z^2) \, dx dz$, and in the thin-film approximation $u_z$ is the only term retained (i.e.~dissipative terms with $u_x$, $v_x$ and $v_z$ are small) so that this part of the dissipation is $\int \frac{\eta}{2} u_z^2 \, dx dz$, as seen in Eq.~\eqref{eq:Phi_appendixB}. We finally note \cite{peschka2018variational} (and some other authors) define the dissipation functional without the prefactor $1/2$, but then incorporate this factor into the definition of $\cal{R}$. Up to notational differences, this (1/2) factor, and the choice of slip law, this directly agrees with Eqs.~(10a) and (10b) of \cite{peschka2018variational}, with the contact line dissipation (10c) there also agreeing with our equation to come in Eq.~\eqref{eq:cldissresultapp}.
}.
The first term in Eq.~\eqref{eq:Phi_appendixB} is the dissipation within the slip region at the surface, with $\lambda$ being the slip length,
and the second term accounts for viscous dissipation from the Navier-Stokes equations under the thin-film approximation. Under this same approximation we can find the velocity $u(z,t)$ from the momentum balance $p_x=\eta u_{zz}$ subject to no tangential shear at the free surface, $u_z|_{z=h}=0$, and our chosen slip condition $u|_{z=0} = (\lambda^2/h)u_z|_{z=0}$, yielding the following (plane Poiseuille film flow with slip) quadratic form
\begin{align}\label{eq:v}
    u(z,t) = \frac{1}{\eta}\frac{\partial p}{\partial x}\left( \frac{z^2}{2} - hz - \lambda^2\right),
\end{align}
where the pressure $p=p(x,t)=\delta F/\delta h$ [see Eqs.~\eqref{EL} and \eqref{dFdh}].
Substituting \eqref{eq:v} into \eqref{eq:Phi_appendixB}, we obtain
\begin{align}
    \Phi&=\frac{1}{2}\int \left(\frac{\partial p}{\partial x}\right)^2\frac{1}{\eta}\left[h\lambda^2+\int_0^h \left(z-h\right)^2 dz\right]dx\nonumber \\
    &= \frac{1}{2}\int \left(\frac{\partial p}{\partial x}\right)^2M_v \, dx,\label{eq:Phi_appendixB_sub1}
\end{align}
where the mobility $M_v(h)=(h^3+3h\lambda^2)/(3\eta)$, i.e.\ the dimensional version of Eq.~\eqref{eq:mobility}.
Now the total flux, given in Eq.~\eqref{eq:J_def}, is
\begin{align}
    J& = \int_0^h u(z) \, dz \nonumber \\
    &=\frac{1}{\eta}\frac{\partial p}{\partial x} \left[\frac{z^3}{6}-\frac{hz^2}{2}-\lambda^2z\right]_0^h\nonumber \\
    &= -\frac{\partial p}{\partial x} M_v.
\end{align}
When we substitute this result into Eq.~\eqref{eq:Phi_appendixB_sub1}, we obtain
\begin{align}
    \Phi = \frac{1}{2}\int \frac{J^2}{M_v} dx.\label{eq:Phi_appendixB_sub2}
\end{align}
When we generalise the above argument to liquid films varying in both the $x$- and $y$-directions, we obtain the first term in Eq.~\eqref{eq:Phi}.

\subsection*{Contact line dissipation}

We now consider the dissipation at the contact line, in the case of a flat homogeneous substrate.
Essentially the arguments follow as with the main text rather than providing an entirely different reasoning, although seeing the connection is quite subtle.

In the (hypothetical) situation of a single one-dimensional moving contact line where all the dissipation occurs at the contact line, then the Rayleighan $\mathcal{R}=\dot{F}+\Phi_{cl}$, where the dissipation function [see Eq.~\eqref{eq:Phi}, or also Eq.~(3.8) in \cite{qian2006variational} or Eq.~(10) in \cite{peschka2018variational}] is the quadratic
\begin{equation}
    \Phi_{cl}=\frac{\dot{X}^2}{2M_{\bm{\Gamma}}}.
\end{equation}
As mentioned, in principle there may be additional terms that are higher order in the contact line velocity $\dot{X}$, but the above leading order quadratic contribution is often sufficient, especially in the small $\dot{X}$ (linear response regime) limit.
The key thing to note is that for small $\dot{X}$, $\Phi_{cl}$ is quadratic.
In the same limit, the time derivative of the free energy $\dot{F}=-\varepsilon |\dot{X}|$; we explicitly determine the coefficient $\varepsilon$ below.
The OVP $\partial\mathcal{R}/\partial\dot{X}=0$ then gives ${-\operatorname{sgn}(\dot{X})}
\varepsilon+\dot{X}/M_{\bm{\Gamma}}=0$, i.e.~$\epsilon = |\dot{X}|/M_{\bm{\Gamma}}$.
Using this in our expression for $\dot{F}$, we obtain $\dot{F}={-}\varepsilon |\dot{X}|={-}\dot{X}^2/M_{\bm{\Gamma}}$.
In other words, the value of $\dot{F}$ evaluated at the minimising value of the contact line velocity $\dot{X}$ is double (and negative of) the corresponding value of $\Phi_{cl}$.
Or, running the same argument in the opposite direction, gives $\Phi_{cl}={-}\dot{F}/2$, when both are evaluated at the value of $\dot{X}$ that minimises $\cal{R}$ \citep{giga2017variational}.

We now move on to consider the mechanism by which dissipation occurs at the contact line during its motion (i.e.\ we determine the coefficient $\varepsilon$ in $\dot{F}$ above).
In our (slip) model, the interface moves at the same velocity as the fluid velocity, i.e.~the contact line itself is a stagnation point in the moving frame of reference.
Thus all dissipation can be considered as coming from the removal/creation of interfaces, which in the case of spreading (receding) is the replacement of dry (wet) substrate surface with a wet (dry) surface and a change in the length of the liquid-vapour free interface. Dissipation is the rate of the decrease in the free energy, so we consider a small advancement of the contact line $\Delta X$ in time $\Delta t$ that changes the free energy by $\Delta F$---see Eq.~\eqref{eq:Delta_F}. Hence the rate of energy dissipation at the contact line is
\begin{align}
\label{eq:A6}
 \Phi_{cl} \approx {-}\frac12\frac{\Delta F}{\Delta t} = {-}\frac{1}{2\Delta t}\left((\gamma_{SL}-\gamma_{SG})\Delta X + \gamma_{LG}(\Delta X\cos\theta)\right),
\end{align}
where it is assumed that in this small time (and small movement $\Delta X$ along the substrate) the free surface has not bent appreciably away from a wedge shape and hence the length of the portion of new interface forms the adjacent side of a right-angled triangle of angle $\theta$ and hypotenuse $\Delta X$.
When there is no motion, $\Delta F = 0$ and $\theta=\theta_s$, so
\begin{align}
    0 = (\gamma_{SL}-\gamma_{SG}) + \gamma_{LG}\cos\theta_s,
\end{align}
which is a rearrangement of the Young equation.
Using this, \eqref{eq:A6} becomes
\begin{align}
\Phi_{cl}=
-\gamma_{LG}\frac{\Delta X}{2\Delta t} (\cos\theta - \cos\theta_s),
\end{align}
which means in the limit $\Delta t \to 0$ we have
\begin{align}\label{eq:B9}
\Phi_{cl} =  -\frac12\dot{X}\gamma_{LG}(\cos\theta - \cos\theta_s).
\end{align}
The term in the brackets above is also proportional to the contact line velocity [see also Eq.~\eqref{eq:V_magnitude_2}],
\begin{align}\label{eq:B10}
    (\cos\theta - \cos\theta_s) = -\frac{\dot{X}}{M_{\bm{\Gamma}} \gamma_{LG}},
\end{align}
where the additional factor $\gamma_{LG}$ in the above compared to \eqref{eq:V_magnitude_2} is due to the fact that we have worked here with dimensional quantities, not the nondimensionalsed ones.
Substituting \eqref{eq:B10} into \eqref{eq:B9} and recalling also that dissipation at the contact line in 2D means dissipation at the sum of the left and right contact points, we have \citep{peschka2018variational, wang2021onsager}
\begin{align}
    \Phi_{cl} = \frac{\dot{X_+}^2 + \dot{X_-}^2}{2 M_{\bm{\Gamma}}}.
    \label{eq:cldissresultapp}
\end{align}
This is the 2D analogue of the second term in Eq.~\eqref{eq:Phi}.

\section{Calculus of variations}
\label{app:cv}

The extension of the calculus of variations for variable domains can be formulated in several ways. In this work, we follow \cite{gelfand1963calculus}, who calculated the variation of a functional over a variable domain as follows. Let $G[h,\Gamma] \equiv \int_\Gamma g(\mathbf{r},h,\nabla h)\,dA$ be a functional of the scalar function $h(\mathbf{r})$ over the simply connected domain $\Gamma \subset \mathbb{R}^2$, and let $h$ vanish at the domain boundary: $h|_{\partial \Gamma}= 0$. We vary both the domain and the function, and the variations will be expressed in terms of the arbitrary vector field $\mathbf{d}: \Gamma \to \mathbb{R}^2$ and scalar field $\psi: \Gamma \to \mathbb{R}$, respectively. Let now $h^*(\mathbf{r}^*)$ denote the varied function defined over the varied domain, where the varied independent and dependent variables are constructed as follows:  
\begin{eqnarray}
\label{rvar} \mathbf{r}^* &\equiv& \mathbf{r} + \epsilon\,\mathbf{d}(\mathbf{r}) + O(\epsilon^2) \\
\label{hvar} h^* &\equiv& h(\mathbf{r}) + \epsilon\,\psi(\mathbf{r}) + O(\epsilon^2) \,,
\end{eqnarray}
where $\epsilon \in \mathbb{R}$ is a small parameter. While Eq.~(\ref{hvar}) is responsible for the variation of $h$, Eq.~(\ref{rvar}) describes the variation of $\Gamma$.
Indeed, for $\mathbf{d}(\mathbf{r})=\mathbf{0}$, $h^*(\mathbf{r}^*)=h(\mathbf{r})+\delta h(\mathbf{r})$ is just the usual variation over a fixed domain with $\delta h(\mathbf{r})=\epsilon\,\psi(\mathbf{r})$. In the general case, the variation of $G$ is defined as: 
\begin{equation}
\delta G \equiv G[h^*,\Gamma^*]-G[h,\Gamma] \,,
\end{equation}
where $\Gamma^* = \{ \mathbf{r}^*(\mathbf{r}): \mathbf{r} \in \Gamma\}$ is the varied domain defined by Eq.~(\ref{rvar}), and $h^* : \Gamma^* \to \mathbb{R}$ is the varied function defined by Eq.~(\ref{hvar}). After lengthy but straightforward algebraic manipulations, $\delta G$ can be expressed in terms of $\epsilon$, $\mathbf{d}(\mathbf{r})$ and $\psi(\mathbf{r})$ as (for more details, see Equations (69)--(95) of Section 37 on pages 168-173 in \cite{gelfand1963calculus}):
\begin{equation}
\label{GFgeneral}
\frac{\delta G}{\epsilon} = \int_{\Gamma} \left\{{\bar{\psi}\,\frac{\delta G}{\delta h}}\right\}dA + \oint_{\partial\Gamma} \left\{ g\,\mathbf{d} + \frac{\partial g}{\partial \nabla h} \,\bar{\psi} \right\} \cdot d\mathbf{l}_\perp + O(\epsilon) \,, 
\end{equation}
where $\frac{\delta G}{\delta h} = g_h - \nabla \cdot \frac{\partial g}{\partial\nabla h}$ is the first functional derivative of $G$ with respect to $h$, $\partial\Gamma$ denotes the boundary of $\Gamma$, $d\mathbf{l}_\perp$ is the outward normal boundary element vector to $\partial\Gamma$, and $\bar{\psi} = \psi - \nabla h \cdot \mathbf{d}$. Due to the terminal boundary condition, $h^*(\mathbf{r}^*)|_{\partial\Gamma^*}=0$ must hold, which, when applied in Eqs.~(\ref{rvar}) and (\ref{hvar}) result in $\psi(\mathbf{r})|_{\partial\Gamma}=0$. Using this in Eq.~(\ref{GFgeneral}) results in:
\begin{equation}
\label{GFterminal}
\frac{\delta G}{\epsilon} = \int_{\Gamma} \left\{ \bar{\psi}\,\frac{\delta G}{\delta h}\right\}\,dA + \oint_{\partial\Gamma} \left\{ \mathbf{d} \cdot \mathbb{B} \right\} \cdot d\mathbf{l}_\perp + O(\epsilon) \,,
\end{equation}
where $\mathbb{B} = g\,\mathbb{I} - \nabla h \otimes \frac{\partial g}{\partial\nabla h}$.

Since $\Gamma \subset \mathbb{R}^2$, with $\partial\Gamma$ being a simple, closed plane curve, $\partial\Gamma$ can be expressed in parametric form as:
\begin{equation}
\label{parcurve}
\mathbf{R}(\chi) = \left( X(\chi), Y(\chi) \right)\,,
\end{equation}
where $\chi \in [0,1]$, $X(0)=X(1)$, and $Y(0)=Y(1)$. $\partial\Gamma^*$ can be obtained by substituting Eq.~(\ref{parcurve}) into Eq.~(\ref{rvar}): $\mathbf{R}^*(\chi) = \mathbf{R}(\chi) + \epsilon\,\mathbf{d}(\mathbf{R}(\chi)) + O(\epsilon^2)$. If the parametrisation of the domain boundary follows the right-hand rule (i.e., increasing $\chi$ corresponds to the counter-clockwise direction along the curve), the outward normal boundary element vector to $\partial\Gamma$ reads as: $d\mathbf{l}_\perp = \mathbf{n}_\perp\,d\chi$, where $\mathbf{n}_\perp=\left( Y_\chi(\chi), -X_\chi(\chi) \right)$, and the subscript $\chi$ stands for derivative with respect to $\chi$. Using these in Eq.~(\ref{GFterminal}), and omitting the higher-order terms in $\epsilon$ result in:
\begin{equation}
\label{GFparametric}
\frac{\delta G}{\epsilon} = \int_{\Gamma} \left\{\bar{\psi}\,\frac{\delta G}{\delta h}\right\}\,dA + \int_0^1 \left\{ \mathbf{d}(\mathbf{R}(\chi)) \cdot \mathbb{T} \cdot \mathbf{n}_\perp \right\}\,d\chi \,,
\end{equation}
where
\begin{equation}
\mathbb{T}=\mathbb{B}|_{\partial\Gamma} = \left( g\,\mathbb{I} - \nabla h \otimes \frac{\partial g}{\partial \nabla h} \right)_{\mathbf{r}=\mathbf{R}(\chi)}    
\end{equation}
is the so-called boundary matrix. 

Let now $h(\mathbf{r},t)$ and $\mathbf{R}(\chi,t)$ be the extension of $h(\mathbf{r})$ and $\mathbf{R}(\chi)$, respectively, where $t \in [0,\infty)$ is an additional independent variable (usually associated with time). The derivative of $G$ with respect to $t$ can be calculated by using Eqs.~(\ref{rvar}), (\ref{hvar}), and (\ref{GFparametric}), as follows. Setting $\epsilon \equiv \Delta t$ (small time increment) and replacing $\mathbf{d}(\mathbf{r})$ and $\psi(\mathbf{r})$ by their respective extensions $\mathbf{d}(\mathbf{r},t)$ and $\psi(\mathbf{r},t)$ provide generic formulae for the change of the liquid thickness and the domain boundary in $\Delta t$ time for an arbitrary dynamics. Consequently, these fields can be related to the partial derivatives of $\mathbf{R}(\chi,t)$ and $h(\mathbf{r},t)$ with respect to $t$ as follows:
\begin{equation}
\label{dRdt}
\begin{split}
\dot{\mathbf{R}} = & \lim_{\Delta t \to 0} \frac{\mathbf{R}(\chi,t+\Delta t)-\mathbf{R}(\chi,t)}{\Delta t} =  \\
= & \lim_{\Delta t \to 0} \frac{\mathbf{R}^*(\chi,t)-\mathbf{R}(\chi,t)}{\Delta t} = \mathbf{d}(\mathbf{R}(\chi,t),t) \,.
\end{split}
\end{equation}
Similarly, the partial derivative of $h(\mathbf{r},t)$ with respect to $t$ (at fixed $\mathbf{r}$) can be written as:
\begin{equation}
\label{dhdt}
\begin{split}
\dot{h} = & \lim_{\Delta t \to 0} \frac{h(\mathbf{r},t+\Delta t)-h(\mathbf{r},t)}{\Delta t} \\
= & \lim_{\Delta t \to 0} \frac{h^*(\mathbf{r},t)-h(\mathbf{r},t)}{\Delta t} = \bar{\psi}(\mathbf{r},t) \,,
\end{split}
\end{equation}
where we used $h^*(\mathbf{r},t)=h^*(\mathbf{r}^*,t)+\nabla h^*|_{\mathbf{r}^*}\cdot(\mathbf{r}-\mathbf{r}^*)+O(\Delta t^2)$, together with $\mathbf{r}-\mathbf{r}^* = -\Delta t\,\mathbf{d}(\mathbf{r},t) + O(\Delta t^2)$ and $\lim_{\Delta t \to 0}\nabla h^*|_{\mathbf{r}^*} = \nabla h|_\mathbf{r}$. Eq.~(\ref{GFparametric}) gives the change of $G$ as a response to the change of the contact line and the liquid thickness. Consequently, using Eqs.~(\ref{dRdt}) and (\ref{dhdt}) in Eq.~(\ref{GFparametric}), then taking the $\Delta t \to 0$ limit yields:
\begin{equation}
\label{dGdt}
\dot{G} = \int_{\Gamma} \left\{\dot{h}\,\frac{\delta G}{\delta h}\right\}\,dA + \int_0^1 \left\{ \dot{\mathbf{R}} \cdot \mathbb{T} \cdot \mathbf{n}_\perp \right\}\,d\chi \,,
\end{equation}
which is the result used in Eq.~\eqref{dFdt}.

\section{Dynamic contact line condition for Section \ref{sec:liq_ridge}}
\label{app:dbc}

In Section \ref{sec:liq_ridge} the dynamic contact angle condition is given in Eq.~\eqref{eq:eg1dbc}. The intermediate steps from Eq.~\eqref{DBCh} are relatively straightforward but numerous so are given below.

Writing Eq.~\eqref{DBCh} in the geometry of Section \ref{sec:liq_ridge} where $x$ is the only spatial coordinate and with free energy from Eq.~\eqref{eq:eg1fullF} gives
\begin{align}
      h_x\left(\lambda^2\frac{\partial^2}{\partial x^2}\left(\frac{-h_x}{\sqrt{1+h_x^2}}\right)\pm\dot{X}_{\pm}
    \right) = 0,
\end{align}
where all quantities are evaluated at the contact points $x=X_\pm(t)$, and continue for the remainder of this Appendix. Expanding the curvature term and rearranging leads to
\begin{align}
\frac{\dot{X}_{\pm}}{\lambda^2} = (1+h_x^2)^{-3/2}\left(
h_{xxx} - 3h_{xx}^2\left(\frac{h_x}{1+h_x^2}\right)\right).
\end{align}
Since 
\begin{align}
    \sin\theta_{\pm} = \mp\frac{h_x}{\sqrt{1+h_x^2}}, \quad
    \cos\theta_{\pm} = \frac{1}{\sqrt{1+h_x^2}},
\end{align}
then 
\begin{align}
    \sin\theta_{\pm}\cos\theta_{\pm} = \frac{\sin(2\theta_{\pm})}{2} =\mp\frac{h_x}{1+h_x^2},
\end{align}
and we find the result that
\begin{equation}
    \frac{\dot{X}_{\pm}}{\lambda^2} = \cos^3\theta_\pm\, \left[ h_{xxx} \pm\,\frac{3}{2}\,\sin(2\,\theta_\pm)\,h_{xx}^2 \right].
\end{equation}
\section{Numerical method}

\subsection{Liquid ridge}
\label{app:num_ridge}

Equations (\ref{hdynridge}), (\ref{bdynridge}), and (\ref{DBCridge}) describe a volume-conserving dynamics on a varying one-dimensional domain. Consequently, the standard method to obtain the numerical solution of the equations is the finite volume method. As the support of $h(x,t)$ is $x \in [X_-(t),X_+(t)]$, where limits vary with time, first we introduce the new independent variable $u \in [0,1]$ satisfying
\begin{equation}
    x = X_-(t) + L(t)\,u \,,
\end{equation}
where $L(t)=X_+(t)-X_-(t)$ is the length of the interval. The new dependent variable is defined as
\begin{equation}
\label{CoVr}\hat{h}(u,t) \equiv \frac{h(X_-(t)+L(t)\,u,t)}{L(t)} \,,
\end{equation}
while the scaled solid surface height reads as $\hat{\phi}(u,t)\equiv \phi(X_-+L\,u)/L$. Differentiating Eq.~(\ref{CoVr}) with respect to $t$ yields (for the sake of readability, we omit the arguments $u$ and $t$ here):
\begin{equation}
\label{hdynur} \dot{\hat{h}} = \frac{\hat{J}_u + \dot{X}_-\,\hat{h}_u + \dot{L}(\hat{h}_u\,u - \hat{h})}{L} \,,
\end{equation}
where $\dot{L}(t)=\dot{X}_+(t)-\dot{X}_-(t)$, $\dot{X}_\pm(t)$ are defined by Eq.~(\ref{bdynridge}). Using Eq.~(\ref{CoVr}) in Eq.~(\ref{hdynridge}) yields  $\dot{h} = \hat{J}_u$, with
\begin{equation}
\hat{J} = \hat{M}_v\,\frac{\partial}{\partial u}\left[ \hat{B}\,\hat{\xi} - \frac{\partial}{\partial u}\left( \frac{\hat{\xi}_u}{\sqrt{1+\hat{\xi}_u^2}} \right) \right] \,,
\end{equation}
where $\hat{\xi}(u,t)=\hat{\phi}(u,t)+\hat{h}(u,t)$, $\hat{B}(t)=B\,L^2(t)$ is the scaled Bond number, and
\begin{equation}
\hat{M}_v(\hat{h})=\frac{M_v(L\,\hat{h})}{L^3} = M_0\,\hat{h}\,(\hat{\lambda}^2+\hat{h}^2) \,,
\end{equation}
is the scaled mobility, where $\hat{\lambda}(t)=\lambda/L(t)$ is the scaled slip length. As $\hat{h}_u=h_x$ and $\hat{\phi}_u = \phi_x$ under Eq.~(\ref{CoVr}), the contact angles simply read as $\theta_\pm(t) = \beta_\pm(t) - \alpha_\pm(t)$, where
\begin{eqnarray}
\mp\,\tan(\alpha_\pm(t)) &=& (\hat{\phi}_u)_{u=0,1} \\ \mp\,\tan(\beta_\pm(t)) &=& (\hat{\xi}_u)_{u=0,1} \,.
\end{eqnarray}
Consequently, Eq.~(\ref{bdynridge}) remains formally unaffected:
\begin{equation}
\label{hdynridget}
\dot{X}_{\pm} = \mp\,M_{\bm{\Gamma}}\,\cos(\alpha_\pm)\,[\cos(\theta_\pm)-\PP] \,.
\end{equation}
Finally, the dynamic boundary condition also can be expressed in terms of the new variables, and reads:
\begin{equation}
\label{DBCnum} \frac{\dot{X}_\pm}{\hat{\lambda}^2} = \cos^3(\beta_\pm)\,\left[ \hat{\xi}_{uuu} \pm \frac{3}{2}\,\sin(2\,\beta_\pm)\,\hat{\xi}^2_{uu} \right] - \hat{B}\,\hat{\xi}_u \,,
\end{equation}
where the boundary dynamics is defined by Eq.~(\ref{hdynridget}), and the derivatives $\hat{\xi}_u$, $\hat{\xi}_{uu}$, and $\hat{\xi}_{uuu}$ are evaluated at $u=0$ and $u=1$, respectively. 

The next step is to express volume conservation in terms of the new variables. Accordingly, $\dot{V} = \frac{d}{dt} \int_0^1 \{ L^2\,\hat{h} \}\,du = 0$, which indicates that the locally conserved quantity is $I\equiv L^2\,\hat{h}$. The quantity $y(x,t)$ is called locally conserved if it satisfies $\dot{y} = J_x$ for some flux $J(t,x,y,y_x,\dots)$. Indeed, taking the derivative of $I$ with respect to $t$, then using Eq.~(\ref{hdynur}), we obtain
\begin{equation}
\label{lconsr} (L^2\,\hat{h})_t = \left\{ \left[ \hat{J} + (\dot{X}_- + \dot{L}\,u)\,\hat{h} \right]\,L \right\}_u \,.
\end{equation}
Since the locally conserved quantity is $L^2\,\hat{h}$, it is more beneficial to proceed with the finite-volume-based semi-discretisation of Eq.~(\ref{lconsr}), rather than the finite difference based semi-discretisation of Eq.~(\ref{hdynur}). Consider now the computational grid $u_i \equiv (i-1)\,\Delta$, where $i=2,\dots,N-1$ (i.e., the number of grid points is $N-2$), and $\Delta=1/(N-1)$ is the grid spacing. As the terminal boundary condition simply translates to $\hat{h}(0,t)=0$ and $\hat{h}(1,t)=0$, $u_1=0$ and $u_N = 1$ are not the part of the grid. Integrating Eq.~(\ref{lconsr}) with respect to $u$ from $u_- \equiv u_i - \frac{\Delta}{2}$ to $u_+ \equiv u_i + \frac{\Delta}{2}$, then using the central value approximation yields the following semi-discretised equations:
\begin{equation}
\label{numerar}\dot{H}_i = \frac{1}{L}\left( \frac{[\hat{J}+(\dot{X}_-+\dot{L}\,u)\,\hat{h}]_{u_-}^{u^+}}{\Delta} - 2\,\dot{L}\,H_i\right) + O(\Delta^2) \,,
\end{equation}
where $H_i(t) = \hat{h}(u_i,t)$ (the value of the analytical solution at $u=u_i$), and $i=2,3,\dots,N-1$. Eq.~(\ref{numerar}) is a second-order accurate discretisation of Eq.~(\ref{hdynur}) for any $O(\Delta^3)$ accurate approximation of the trace $[\hat{J}+(\dot{X}_-+\dot{L}\,u)\,\hat{h}]_{u_-}^{u^+}$. Suitable schemes of the lowest order read as:
\begin{eqnarray}
\frac{[\hat{h}]_{u_-}^{u^+}}{\Delta} &=& \frac{H_{i+1}-H_{i-1}}{2\,\Delta} + O(\Delta^2)\,, \\
\frac{[\hat{h}\,u]_{u_-}^{u^+}}{\Delta} &=& \frac{H_{i+1}-H_{i-1}}{2\,\Delta}\,u_i + H_i + O(\Delta^2)\,, \\
\frac{[\hat{J}]_{u_-}^{u^+}}{\Delta} &=& \frac{\hat{J}^+_i - \hat{J}^-_i}{\Delta} + O(\Delta^2) \,,
\end{eqnarray}
where the fluxes read as 
\begin{eqnarray}
\hat{J}_i^+ &=& \hat{M}_v\left( \frac{H_i+H_{i+1}}{2} \right)\,\frac{\hat{\omega}_{i+1}-\hat{\omega}_i}{\Delta}\,, \\
\hat{J}^-_i &=& \hat{M}_v\left( \frac{H_{i-1}+H_i}{2} \right)\,\frac{\hat{\omega}_i-\hat{\omega}_{i-1}}{\Delta} \,,
\end{eqnarray}
and the functional derivative is approximated as
\begin{equation}
\hat{\omega}_i = \hat{B}\,\hat{\xi}_i - \frac{\hat{\xi}_{uu,i}}{(1+\hat{\xi}_{u,i}^2)^{3/2}} \,,
\end{equation}
where $\hat{\xi}_i(t)=\hat{\phi}_i(t)+H_i(t)$, $\hat{\xi}_{u,i}(t)=\hat{\phi}_{u,i}(t)+\hat{h}_{u,i}(t)$, and $\hat{\xi}_{uu,i}(t)=\hat{\phi}_{uu,i}(t)+\hat{h}_{uu,i}(t)$. While $\hat{\phi}_i(t) = \hat{\phi}(u_i,t)$, $\hat{\phi}_{u,i}(t) = \hat{\phi}_u(u_i,t)$ and $\hat{\phi}_{uu,i}(t) = \hat{\phi}_{uu}(u_i,t)$ are analytical as long as $\phi(x)$ is analytical, the derivatives of the liquid height are approximated by using central differences: $\hat{h}_{u,i} = \frac{H_{i+1}-H_{i-1}}{2\,\Delta}$ and $\hat{h}_{uu,i} = \frac{H_{i-1}+H_{i+1}-2\,H_i}{\Delta^2}$. The remaining step is to apply the boundary conditions. Obviously, $H_1=H_N=0$ from the terminal boundary condition, but this is not sufficient for Eq.~(\ref{numerar}) for $i=2$ and $i=N-1$ for the following reason.  Eq.~(\ref{numerar}) can be written in the general form of $\dot{H} = \zeta(H_{i-2},H_{i-1},H_i,H_{i+1},H_{i+2}) + O(\Delta^2)$, i.e., the second-order scheme requires data from the two nearest neighbours on the grid in each direction. While these are available for $i=3,4,\dots,N-2$ due to the terminal boundary condition, $\phi_{0}$ $H_{N+1}$ enter the equation for $i=2$ and $i=N-1$, respectively. As $u=-\Delta$ and $u=1+\Delta$ are not the part of the support of $\hat{h}(u,t)$, $\hat{h}(u,t)$ must be extended and $H_0$ and $H_{N+1}$ must be approximated. Firstly, Eq.~(\ref{numerar}) can be written as $\dot{H}_2 = a_2\,\frac{H_0}{\Delta^4} + \dots + O(\Delta^2)$ and $\dot{H}_{N-1} = a_{N-1}\,\frac{H_{N+1}}{\Delta^4} + \dots + O(\Delta^2)$. Therefore, to retain the $O(\Delta^2)$ accuracy of the scheme, $H_0$ and $H_{N+1}$ must be approximated with accuracy $O(\Delta^6)$, which can be done by discretising the dynamic boundary condition with accuracy $O(\Delta^6)$. Accordingly, we consider the generic schemes
\begin{eqnarray}
\label{BDCd1r}
\hat{h}_u|_{u=1,0} &=& \Delta^{-1} \left(\mathbf{a}^\pm_1 \cdot\mathbf{H}+b^\pm_1\,H_\pm\right)+O(\Delta^6)\,, \\
\label{BDCd2r} \hat{h}_{uu}|_{u=1,0} &=& \Delta^{-2} \left(\mathbf{a}^\pm_2 \cdot\mathbf{H}+b^\pm_2\,H_\pm\right)+O(\Delta^6)\,, \\
\label{BDCd3r} \hat{h}_{uuu}|_{u=1,0} &=& \Delta^{-3} \left(\mathbf{a}^\pm_3 \cdot\mathbf{H}+b^\pm_3\,H_\pm\right)+O(\Delta^6) \,,
\end{eqnarray}
where $H_\pm=H_{N+1},H_0$, respectively, $\mathbf{H}=(H_2,H_3,\dots,H_{N-1})$, $\mathbf{a}^\pm_m$ is an $(N-2)$-component constant vector and $b^{(m)}_\pm$ is a constant ($m=1,2,3$). Using Eqs.~(\ref{BDCd1r})-(\ref{BDCd3r}) in Eq.~(\ref{DBCnum}), the dynamic boundary condition can be written in the following general form:
\begin{equation}
\label{DBCn} q_\pm(\mathbf{H},H_\pm)+O(\Delta^6) = 0\,.
\end{equation} 
Let $\bar{H}_\pm$ denote the solution of the equation $q_\pm(\mathbf{H},\bar{H}_\pm)=0$, which can be obtained numerically by using Newton iteration. As $\bar{H}_\pm$ approximates $H_\pm$ with accuracy $O(\Delta^6)$, $H_\pm$ can be replaced by $\hat{H}_\pm$ in Eq.~(\ref{numerar}) without affecting the order of accuracy. Finally, suitable choices for the weight vectors of the lowest order read as: 
\begin{eqnarray*}
\mathbf{a}_1^+ &=& \left(0,0,\dots,0, -\frac{1}{30},\frac{1}{4},
-\frac{5}{6},
\frac{5}{3},-\frac{5}{2} \right) \\
\mathbf{a}_2^+ &=& \left(0,0,\dots,0, \frac{11}{180},-\frac{1}{2},\frac{9}{5},-\frac{67}{18},\frac{19}{4},-\frac{27}{10} \right) \\
\mathbf{a}_3^+ &=& \left(0,0,\dots,0, \frac{1}{48},-\frac{9}{40},\frac{67}{60},-\frac{407}{120},\frac{57}{8},-\frac{269}{24},
\frac{731}{60}
\right) \,,
\end{eqnarray*}
while $b_1^+=\frac{1}{6}$, $b_2^+=\frac{7}{10}$, and $b_3^+=\frac{469}{240}$. $\mathbf{a}_m^-$ coincides with $\mathbf{a}_m^+$ with components of reversed order, and $b_m^-=b_m^+$ for $m=1,2,3$. To accelerate the convergence of the Newton method, the iteration is started from an $O(\Delta^6)$ accurate approximation: $\bar{H}_\pm^{(0)} \equiv \mathbf{a}_0^\pm\cdot\mathbf{H}$, where
\begin{equation}
\mathbf{a}_0^+=\left(0,0,\dots,0,-\frac{1}{6},1,-\frac{5}{2},\frac{10}{3} \right) \,,
\end{equation}
while $\mathbf{a}_0^-$ coincides with $\mathbf{a}_0^+$ with reversed order of the components.

\subsection{Axisymmetric drop}
\label{app:num_axi}

The discretisation strategy of Eqs.~(\ref{axialht}), (\ref{axialRt}) and (\ref{DBCax}) is the same as the one presented in the previous section, and therefore we only highlight the differences here. In the case of axial symmetry, the variable transformation reads as:
\begin{equation}
\label{CoV}\hat{h}(u,t) \equiv \frac{h(R\,u,t)}{R} \,,
\end{equation}
where $u \in [0,1]$ is the new independent variable satisfying
\begin{equation}
    r = R(t)\,u \,.
\end{equation}
The scaled solid height reads as $\hat{\phi}(u,t)\equiv \frac{\phi(R\,u)}{R}$. Differentiating Eq.~(\ref{CoV}) with respect to $t$ yields:
\begin{equation}
\label{hdynu}\dot{\hat{h}} = \frac{\dot{h} + \dot{R}\,(\hat{h}_u\,u-\hat{h})}{R} \,,
\end{equation}
where $\dot{R}$ is defined by Eq.~(\ref{axialRt}). Furthermore, using Eq.~(\ref{CoV}) in Eq.~(\ref{axialht}) yields  $\dot{h} = \frac{1}{u}\,\frac{d}{du}(u\,\hat{J})$, where
\begin{equation}
\hat{J} = \hat{M}_v\,\frac{d}{du}\left[ \hat{B}\,\hat{\xi} - \frac{1}{u}\,\frac{d}{du}\left( \frac{u\,\hat{\xi}_u}{\sqrt{1+\hat{\xi}_u^2}} \right) \right] \,,
\end{equation}
$\hat{B}(t)=B\,R^2(t)$ is the scaled Bond number, and $\hat{M}_v(\hat{h})=\frac{M_v(\hat{h}\,R)}{R^3} = M_0\,\hat{h}\,(\hat{\lambda}^2+\hat{h}^2)$ is the scaled mobility, where $\hat{\lambda}(t)=\lambda/R(t)$ is the scaled slip length.
The boundary conditions also can be expressed in terms of the new variables, and read as: $\hat{h}_{u}|_{u=0} = \hat{h}_{uuu}|_{u=0} = 0$, $\hat{h}(1,t)=0$, while the dynamic boundary condition defined by Eq.~(\ref{DBCax}) reads
\begin{equation}
\label{DBCu} \frac{\dot{R}}{\hat{\lambda}^2} = \sin(\beta) - \hat{B}\,\xi_u + \cos^3(\beta)\,\hat{\omega}(\hat{\xi}_{u},\hat{\xi}_{uu},\hat{\xi}_{uuu}) \,,
\end{equation}
where $\dot{R}$ is defined by Eq.~(\ref{axialRt}), and $\hat{\omega}(\hat{\xi}_{u},\hat{\xi}_{uu},\hat{\xi}_{uuu}) = \hat{\xi}_{uuu}+\frac{3}{2}\,\sin(2\,\beta)\,\hat{\xi}_{uu}^2+\hat{\xi}_{uu}$. As the volume of the drop is constant, the locally conserved quantity is $I\equiv R^3\,\hat{h}\,u$, which satisfies the following equation:
\begin{equation}
\label{lcons}(R^3\,\hat{h}\,u)_t =  \left[ R^2\,\left(\dot{R}\,\hat{h}\,\xi^2+\hat{J}\,u\right) \right]_u \,.
\end{equation}
Consider now the computational grid $u_i \equiv (i-1/2)\,\Delta$, where $i=1,2,\dots,N-1$, and $\Delta=1/(N-1/2)$ is the grid spacing.
With this choice, $u_1=\Delta/2$ and $u_N=1$, and therefore the boundary conditions can be easily applied. The semi-discretised equations read as:
\begin{equation}
\label{numeraax}\dot{H}_i = \frac{1}{R}\left( \frac{[\hat{J}\,u+\dot{R}\,\hat{h}\,u^2]_{u_-}^{u^+}}{u_i\,\Delta} - 3\,\dot{R}\,H_i\right) + O(\Delta^2) \,,
\end{equation}
where $H_i = \hat{h}(u_i,t)$. Suitable schemes of the lowest order to approximate the trace in Eq.~(\ref{numeraax}) read as:
\begin{eqnarray}
\frac{[\hat{h}\,u^2]_{u_-}^{u^+}}{u_i\,\Delta} &=& \frac{H_{i+1}-H_{i-1}}{2\,\Delta}\,u_i + 2\,H_i + O(\Delta^2)\,, \\
\frac{[\hat{J}\,u]_{u_-}^{u^+}}{u_i\,\Delta} &=& \frac{\hat{J}_i^+ - \hat{J}^-_i}{\Delta} + \frac{\hat{J}_i}{u_i} + O(\Delta^2) \,,
\end{eqnarray}
where 
\begin{eqnarray}
\hat{J}_i &=& \hat{M}_v(H_i)\,\frac{\hat{\omega}_{i+1}-\hat{\omega}_{i-1}}{2\,\Delta}\,, \\
\hat{J}^+_i &=& \hat{M}_v\left(\frac{H_i+H_{i+1}}{2}\right)\,\frac{\hat{\omega}_{i+1}-\hat{\omega}_i}{\Delta}\,, \\
\hat{J}^-_i &=& \hat{M}_v\left(\frac{H_{i-1}+H_i}{2}\right)\,\frac{\hat{\omega}_i-\hat{\omega}_{i-1}}{\Delta} \,,
\end{eqnarray}
where
\begin{equation}
\hat{\omega}_i = \hat{B}\,\hat{\xi}_{u,i} - \frac{\hat{\xi}_{u,i}^3+\hat{\xi}_i+u_i\,\hat{\xi}_{uu,i}}{u_i\,\left(1+\hat{\xi}_{u,i}^2\right)^{3/2}} \,,
\end{equation}
where the derivatives of $\hat{h}(u,t)$ are approximated by using centred differences. Finally, the boundary conditions are applied as follows: At $u=0$, $\hat{h}_u=\hat{h}_{uuu}=0$ is naturally satisfied by extending $\hat{h}(u,t)$ as $\hat{h}(u,t)\equiv \hat{h}(-u,t)$ for $u<0$. Accordingly, the first two neighbours of $H_1$ (located at $u_1=\Delta/2$) to the left are $H_2$ and $H_3$ at locations $u_0=-\Delta/2$ and $u_{-1}=-\Delta/2-\Delta$, respectively. At $u=1$, $H_N=0$ applies due to the terminal boundary condition, while $H_{N+1}$ is approximated by obtaining the numerical solution of Eq.~(\ref{DBCu}) with accuracy $O(\Delta^6)$ as described in the previous section.

\bibliographystyle{jfm}

\end{document}